%% file: main.tex
\documentclass[aps, reprint, superscriptaddress, floatfix, showpacs, amsmath, amssymb]{revtex4-1}

\usepackage{graphicx,color}
\usepackage{dcolumn}
\usepackage{bm}
\usepackage{hyperref}
\usepackage{lineno}

\begin{document}

\title{Azimuthal transverse single-spin asymmetries of inclusive jets and identified hadrons within jets from polarized $pp$ collisions at $\sqrt{s}$ = 200 GeV}

\include{star-author-list-2022-09-13.aps}

\date{\today}
\begin{abstract}
The STAR Collaboration reports measurements of the transverse single-spin asymmetries, $A_N$, for inclusive jets and identified `hadrons within jets' production at midrapidity from transversely polarized $pp$ collisions at $\sqrt{s}$ = 200 GeV, based on data recorded in 2012 and 2015. The inclusive jet asymmetry measurements include $A_N$ for inclusive jets and $A_N$ for jets containing a charged pion carrying a momentum fraction $z>0.3$ of the jet momentum. The identified hadron within jet asymmetry measurements include the Collins effect for charged pions, kaons and protons, and the Collins-like effect for charged pions. The measured asymmetries are determined for several distinct kinematic regions, characterized by the jet transverse momentum $p_{T}$ and pseudorapidity $\eta$, as well as the hadron momentum fraction $z$ and momentum transverse to the jet axis $j_{T}$. These results probe higher momentum scales ($Q^{2}$ up to $\sim$\,900 GeV$^{2}$) than current, semi-inclusive deep-inelastic scattering measurements, and they provide new constraints on quark transversity in the proton and enable tests of evolution, universality and factorization breaking in the transverse-momentum-dependent formalism.

\end{abstract}

\pacs{}

\maketitle


\section{Introduction\label{sec:Introduction}}
\subsection{Background}
The nature of Quantum Chromodynamics (QCD) is illuminated by investigations into how the proton mass, charge, and spin manifest from the properties of the more fundamental quarks and gluons. Proton spin structure measurements have proven to be a powerful tool for probing the robustness of theoretical frameworks in QCD. It was the failure of traditional leading twist, collinear perturbative QCD predictions \cite{ref:Kane_1978} to describe a series of transverse single-spin asymmetry ($A_N$) measurements in polarized proton-proton collisions \cite{ref:Dragoset_1978,ref:Bonner_1988,ref:Bonner_1990,ref:Adams_1991a,ref:Adams_1991b,ref:Bravar_1996,ref:Adams_2004} that fueled the rapid and rich development of twist-3 \cite{ref:EfremovTeryaev,ref:QiuSterman, ref:Qiu_1998} and transverse-momentum-dependent (TMD) \cite{ref:Collins_1981,ref:Collins_1985,ref:SiversFunction,ref:CollinsFF} factorization schemes.
The success of these twist-3 and TMD frameworks in describing transverse spin effects has generated worldwide interest and spawned new experimental programs aimed at studying twist-3 and TMD observables in electron-positron, lepton-proton, and proton-proton collisions.

The twist-3 factorization scheme applies to measurements with a single hard scale $Q$ which is much larger than $\Lambda_{QCD}$, such as measurements of $A_N$ in inclusive pion and jet production.  In contrast, the TMD factorization scheme applies to measurements where there is another momentum scale in addition to the hard scale, such as the transverse momentum of an identified final-state hadron. In this case the softer scale can be as small as $\Lambda_{QCD}$ and should be much smaller than $Q$.  The two frameworks are closely related, as twist-3 functions can be written as transverse momentum moments of the related TMD functions \cite{ref:Boer_2003}.  They have also been shown to describe the same physics for many scenarios in the intermediate transverse momentum region where both approaches can be applied. \cite{ref:Ji_2006,ref:Koike_2008,ref:Yuan_2009,ref:Bacchetta_2008}.

Examples of  spin-dependent TMD observables are the Sivers \cite{ref:SiversFunction} and Collins \cite{ref:CollinsFF} asymmetries measured in semi-inclusive deep-inelastic scattering (SIDIS) by the HERMES \cite{ref:HERMES:2004mhh,ref:Airapetian_2009,ref:Airapetian_2010,ref:Airapetian_2020}, COMPASS \cite{ref:Alekseev_2009,ref:Adolph_2015,ref:Adolph_2017} and Jefferson Lab Hall A \cite{ref:Qian_2011,ref:Zhao_2014} experiments, the Collins fragmentation functions extracted from $e^+e^-$ annihilation at BELLE \cite{ref:Seidl_2008,ref:Belle:2019nve}, BaBar \cite{ref:Lees_2014,ref:Lees_2015}, BESIII \cite{ref:Ablikim_2016}, the limit on the Sivers effect in dijet production set by STAR \cite{ref:STAR_Dijet_Sivers07}, the Collins asymmetry for $\pi^0$ in forward rapidity electromagnetic jets measured by STAR \cite{ref:STAR_emJets}, and the $A_N$ of $W^{\pm}$ and $Z$ measured by STAR \cite{ref:Adamczyk_2016}, as well as the Drell-Yan di-muons measured by COMPASS \cite{ref:Aghasyan_2017}.  

The Collins effect represents a particularly interesting case.  In SIDIS, the Collins asymmetry \cite{ref:CollinsFF} arises from the convolution of the TMD transversity parton distribution function (PDF), $h_1^q\left(x,k_T,Q^2\right)$, and the TMD Collins fragmentation function (FF), $H^{\perp}_{1\,\pi/q}\left(z,\kappa_T,Q^2\right)$.  The TMD transversity distribution  describes the transverse polarization of quarks in a transversely polarized proton as a function of their longitudinal momentum fraction ($x$), and transverse momentum ($k_T$).  Transversity is one of the three leading-twist PDFs of the nucleon that survive integration over parton transverse momentum.  However, unlike the unpolarized and helicity distributions, transversity is chiral-odd, so much less is known about it because it is quite challenging to extract via inclusive deep-inelastic scattering~\cite{ref:Accardi:2017pmi}.  Nonetheless, it is essential for a complete understanding of nucleon structure.  Lattice QCD calculations of the transversity distribution have been performed \cite{ref:XJi_2013, ref:Chen_2016, ref:Alexandrou_2018, ref:Constantinou_2021} and even incorporated into global analyses of the world datasets~\cite{ref:Lin_2018}. The integral of the transversity distribution gives the nucleon tensor charge, which plays a key role in low-energy searches for physics beyond the standard model involving, for example, the neutron electric dipole moment \cite{ref:Bhattacharya_2016} and $\beta$-decay \cite{ref:Dubbers_2011}.  Furthermore, the difference between the helicity and transversity distributions has been suggested to provide a direct, $x$-dependent connection to quark orbital angular momentum \cite{ref:sivers2011adventure}.

The Collins FF describes the azimuthal distribution of hadron fragments emitted from a transversely polarized quark as a function of the fraction of the quark momentum carried by the hadron ($z$), and the hadron momentum transverse to the quark direction ($\kappa_T$).  This function provides an excellent testing ground to investigate fundamental properties of TMDs, including factorization, universality and evolution.   The Collins effect has been shown to be universal in $e^+e^-$ collisions and SIDIS \cite{ref:Metz_2002,ref:Collins_2004, ref:Meissner_2009}.  Several groups have performed global fits of the existing Collins asymmetry data to extract transversity and the Collins FF \cite{ref:Anselmino_2015,ref:Kang_2016,ref:D_Alesio_2020,ref:Cammarota2020origin}.  Transversity has also been extracted from global analyses~\cite{ref:Radici_2018,ref:Benel2019constrained} of di-hadron interference fragmentation function measurements \cite{ref:Collins_1994,ref:Bacchetta_2004} in $e^+e^-$, SIDIS and $pp$ collision data \cite{ref:HERMES:2008mcr,ref:COMPASS:2012bfl,ref:COMPASS:2014ysd,ref:Belle:2011cur,ref:STAR:2015jkc,ref:Belle:2017rwm}.

Whether the universality of TMD PDFs and FFs can be extended to $pp$ collisions is still an open question. On one hand, general arguments have shown that factorization is violated in hadron-hadron collisions for TMD PDFs like the Sivers function \cite{ref:Collins_2007,ref:Rogers_2010}.  On the other hand, explicit calculations \cite{ref:Yuan2007,ref:Yuan2008} have shown that the eikonal propagators that violate factorization of the Sivers function in hadron-production do not contribute to the Collins effect at the one- or two-gluon exchange level, thereby preserving its universality.

More recently, detailed calculations have investigated the Collins effect in $pp$ collisions using soft-collinear effective theory (SCET) \cite{ref:Kang_2017a,ref:Kang_2017b}.  They find that, in contrast to current measurements of {\em single-hadron} production in SIDIS, if the azimuthal distribution of hadrons is measured about their parent jet axis, the Collins effect both in $pp$ collisions and SIDIS involves a mixture of collinear and TMD factorization. In the case of $pp$ collisions, in the first step, the jet production involves standard collinear factorization including a convolution of the collinear transversity PDF, $h_1^q\left(x_1,Q^2\right)$, for the polarized proton and the unpolarized PDF, $f_q\left(x_2,Q^2\right)$, for the unpolarized proton. In the second step, the scattered transversely polarized quark fragments according to the Collins FF, $H^{\perp}_{1\,\pi/q}\left(z,j_T,Q^2\right)$, where $z$ is the hadron momentum fraction within the jet and $j_T$ is the hadron momentum transverse to the jet axis.  If the parent jet has been reconstructed using the standard jet axis, in this case using the anti-$k_T$ algorithm \cite{ref:AntiKtJets}, Refs.\@ \cite{ref:Kang_2017a,ref:Kang_2017b} show that the Collins FF measured in $pp$ collisions is the same as that in SIDIS. The fact that the collinear transversity distribution enters in jet-plus-hadron measurements, in contrast to the TMD transversity distribution in single-hadron SIDIS, means that the previous are a more direct probe of the Collins FF~\cite{ref:Kang_2017a}. Currently, only $pp$ collisions at the Relativistic Heavy Ion Collider (RHIC) allow such studies while similar measurements using SIDIS jets will require the Electron-Ion Collider~\cite{ref:Kang:2021ffh}.

 \begin{figure}[h]
 \includegraphics[width=\columnwidth]{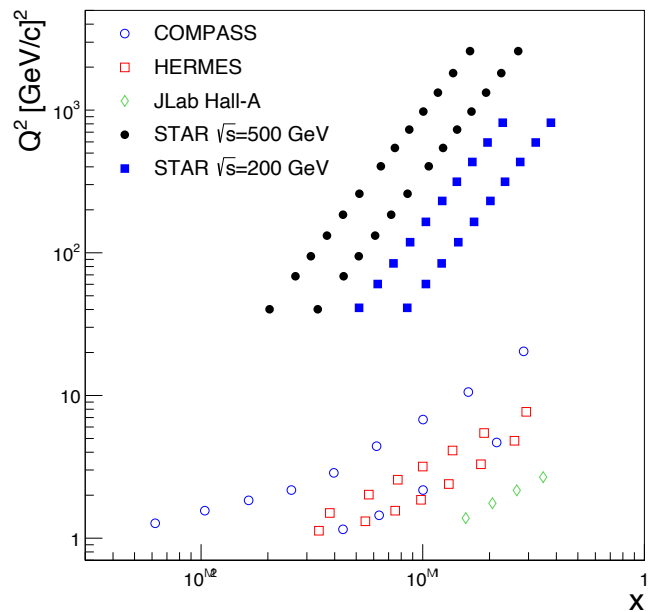}%
 \caption{\label{fig:rhicKine}The kinematic coverage of these results from STAR at $\sqrt{s}$ = 200 GeV compared to the coverage of previous STAR $pp$ measurements at $\sqrt{s}$ = 500 GeV \cite{ref:500GeVCollins} and SIDIS measurements that also target transversity~\cite{ref:Airapetian_2009,ref:Airapetian_2010,ref:Airapetian_2020,ref:Alekseev_2009,ref:Adolph_2015,ref:Adolph_2017,ref:Qian_2011,ref:Zhao_2014}.} 
 \end{figure}

In 2018, the STAR Collaboration reported the first measurements of the Collins effect in $pp$ collisions, based on a small data set at a center-of-mass energy $\sqrt{s}$ = 500 GeV that was recorded in 2011\@ \cite{ref:500GeVCollins}.  The results are consistent with predictions based on global analyses of $e^+e^-$ and SIDIS data \cite{ref:Kang_2017b,ref:D_Alesio_2017}, thereby supporting the expectation that the universality of the Collins effect extends to $pp$ collisions. But, Kang and his collaborators~\cite{ref:Kang_2017b} emphasized that effects that fall outside of standard SCET \@ \cite{ref:Collins_2007,ref:Dasgupta_2001,ref:Banfi_2002} are not included in their calculations. These effects are expected to make at most small contributions to the measured asymmetries because they can only arise at the three-gluon exchange level or higher~\cite{ref:Yuan2008}. Thus, higher precision $pp$ measurements are essential to probe the universality of the Collins FF and search for potential factorization breaking of TMD FFs in $pp$ collisions.

In this paper, the STAR Collaboration presents the first measurements of the Collins effect in $pp$ collisions at $\sqrt{s}$ = 200 GeV.  Figure \ref{fig:rhicKine} shows the kinematic coverage of the current measurements in comparison to those from SIDIS and the previous 500 GeV data. These STAR measurements overlap much of the $x$ range of SIDIS but at a dramatically higher range of $Q^2$. In addition to the current SIDIS results, studying the Collins effect at higher values of $Q^2$ will provide necessary input on the evolution of the TMD functions, a topic of vigorous discussion in the nuclear physics community\cite{ref:Aybat_2012,ref:Echevarria_2014,ref:Kang_2016}. Unlike the collinear case, TMD evolution includes a non-perturbative contribution \cite{ref:Collins_1985} and it cannot be derived from first principles. This effect has been studied in a global analysis of the Collins asymmetries in SIDIS and electron-positron annihilation \cite{ref:Kang_2016} before the STAR hadron in jet data, where the same TMD evolution effect should also be applied~\cite{ref:Kang_2017a}. At the same time, having these asymmetry measurements from $pp$ collisions, as well as SIDIS and $e^+e^-$ experiments, will allow the universality of these functions to be studied and limits on factorization-breaking to be quantified.

\subsection{Azimuthal modulations}
\label{sec:AziMod}
In $pp$ collisions, the Collins effect manifests itself as a spin-dependent azimuthal modulation of hadrons about their parent jet axis \cite{ref:Yuan2007,ref:DAlesio2011}. However, measurement of the azimuthal single-spin asymmetry, $A_{UT}$,  for $\pi^\pm$ inside a jet in $p^\uparrow +p$ collisions opens the door simultaneously to access many additional observables which probe the proton's internal spin and transverse momentum structure. These observables are described in the following expression for the relative difference of the spin-dependent cross sections \cite{ref:DAlesio2011}:
\begin{align}
\label{eq:crosssec}
 &\frac{d\sigma^\uparrow\left(\phi_S,\phi_H\right) -d\sigma^\downarrow\left(\phi_S,\phi_H\right)}{d\sigma^\uparrow\left(\phi_S,\phi_H\right) +d\sigma^\downarrow\left(\phi_S,\phi_H\right)} \nonumber \\ 
 &\propto A_{UT}^{\sin\left(\phi_S\right)}\sin\left(\phi_S\right) \nonumber \\
 &+ A_{UT}^{\sin\left(\phi_S-\phi_H\right)}\sin\left(\phi_S-\phi_H\right) \nonumber\\
 &+A_{UT}^{\sin\left(\phi_S-2\phi_H\right)}\sin\left(\phi_S-2\phi_H\right) \nonumber \\
 &+ A_{UT}^{\sin\left(\phi_S+\phi_H\right)}\sin\left(\phi_S+\phi_H\right) \nonumber\\
 &+A_{UT}^{\sin\left(\phi_S+2\phi_H\right)}\sin\left(\phi_S+2\phi_H\right) .
\end{align}
 
\begin{figure}[h]
 \includegraphics[width=\columnwidth]{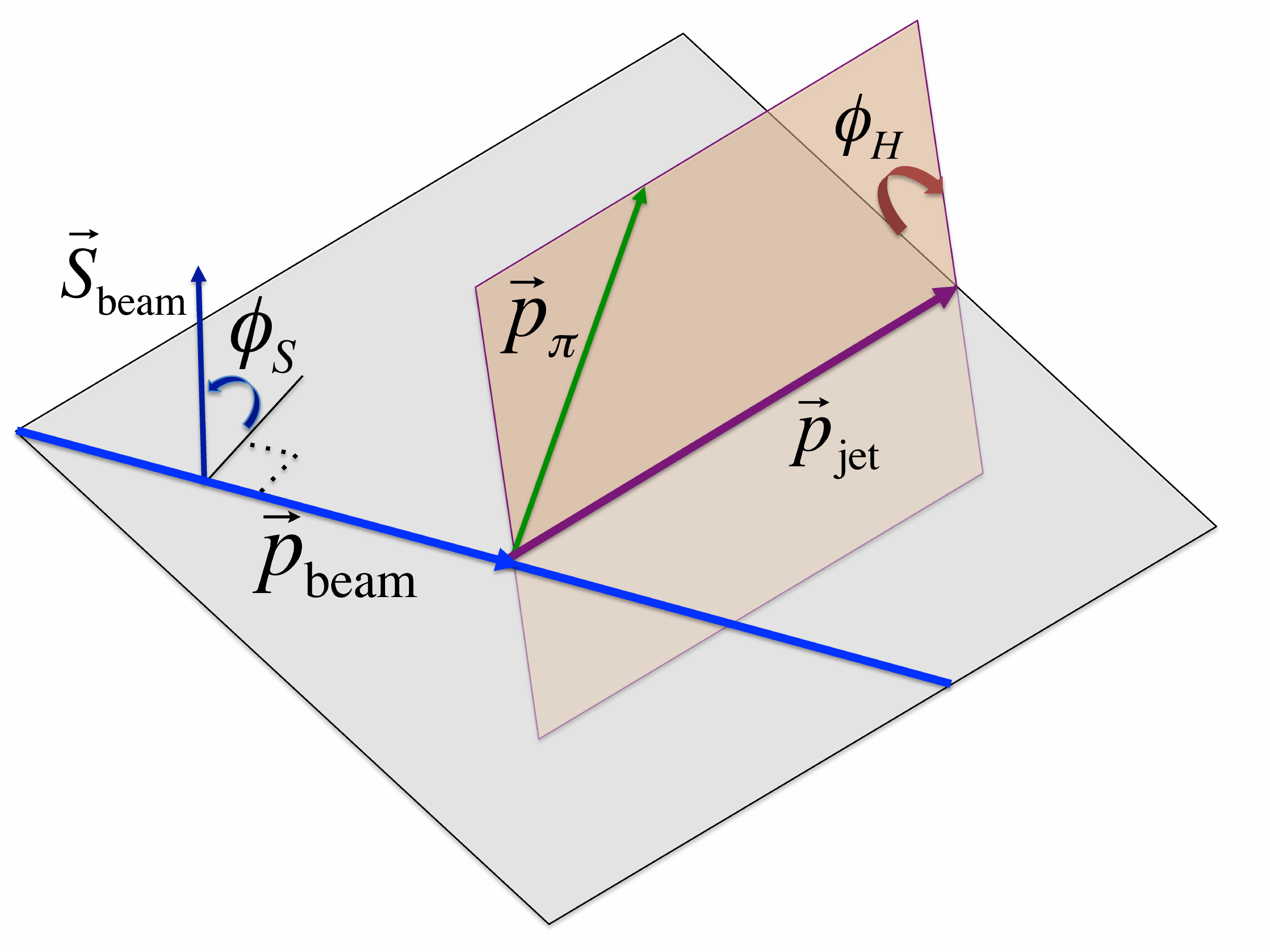}%
 \caption{\label{fig:azimuthalAngles}Representation of the reaction plane and the orientation of the angles $\phi_S$ and $\phi_H$ relative to this plane.}
 \end{figure}

\noindent The $A_{UT}$ coefficients can be expressed as combinations of parton distribution and fragmentation functions. The sine terms depend upon $\phi_S$, the angle between the proton spin direction and the reaction plane, and $\phi_H$, the angle of the pion momentum transverse to the jet axis relative to the reaction plane. In the  $p^\uparrow+p\rightarrow \textrm{jet}+\pi^{\pm}+X$ process, the reaction plane is defined by the incoming polarized beam ($\vec{p}_{\mathrm{beam}}$) and scattered jet momentum ($\vec{p}_{\mathrm{jet}}$). Figure \ref{fig:azimuthalAngles} shows the orientation of these angles with respect to the reaction plane.

In Eq.\@ (\ref{eq:crosssec}), the inclusive jet asymmetry is the coefficient of the $\sin\left(\phi_S\right)$ term.  STAR has measured this asymmetry before at midrapidity in $\sqrt{s}$ = 200 GeV $pp$ collisions \cite{ref:StarSivers2006}, where it was referred to as $A_N$, and also at $\sqrt{s}$ = 500 GeV~\cite{ref:500GeVCollins}. $\mathrm{A_{N}DY}$ has also measured this asymmetry at forward rapidity in $pp$ collisions at $\sqrt{s}$ = 500 GeV~\cite{ref:AnDY:2013att}. This asymmetry is sensitive to the initial state twist-3 quark-gluon correlators, which are described by the Efremov-Teryaev-Qiu-Sterman (ETQS) function \cite{ref:EfremovTeryaev,ref:QiuSterman,ref:Qiu_1998}. These correlators are related to the leading twist TMD Sivers function \cite{ref:SiversFunction,ref:Boer_2003,ref:Ji_2006,ref:Koike_2008} which encapsulates the correlation between $k_T$ and the transverse spin of the parent proton. At low values of jet transverse momentum ($p_T$), the majority of the jets in proton-proton collisions originate from hard-scattered gluons.  Therefore, the inclusive jet asymmetries reported here can give precise input on the twist-3 correlators associated with the gluon Sivers function.  In addition, the asymmetries reported here for jets that contain a high-$z$ charged pion  provide information regarding the twist-3 correlators associated with the quark Sivers functions \cite{ref:Gamberg_2013, ref:D_Alesio_2011, ref:DAlesio2011}.  Recently, PHENIX reported measurements of $A_N$ for midrapidity $\pi^0$ and $\eta$ mesons \cite{ref:PHENIX_pi0_AN} and isolated direct photons \cite{ref:PHENIX_gamma_AN} in 200 GeV $pp$ collisions which also provide information regarding the twist-3 correlators associated with the gluon Sivers function.  However, the $\pi^0$ and $\eta$ $A_N$ can have contributions from final-state effects in addition to the initial-state Sivers effect, whereas the inclusive jet asymmetries reported here are believed to be free of final-state contributions.  The isolated direct photon measurements are also free of final-state contributions, but the statistical precision achieved was limited.

The coefficient of the $\sin\left(\phi_S-\phi_H\right)$ term is the Collins asymmetry. As described above, it is sensitive to the combination of the transversity PDF and the Collins FF. The gluon transversity distribution in the proton has to be zero due to the conservation of angular momentum. As a result, this asymmetry is expected to increase with increasing jet-$p_T$ as the fraction of gluons participating in the hard scattering decreases.  Multi-dimensional separations of the Collins asymmetry, including the dependence on jet-$p_T$ for two different pseudorapidity ($\eta$) regions and the dependence on hadron-$j_T$ for four different hadron-$z$ regions,  are provided to separate the initial-state transversity effects from the kinematics of the Collins FF in the final-state. Collins asymmetries for kaons and protons in jets as complementary probes of the dynamical origins of the Collins FF are also presented in this analysis.

While gluons cannot carry any transverse spin in the proton, they can be linearly polarized.  The coefficient of the $\sin\left(\phi_S-2\phi_H\right)$ term, the so called `Collins-like' asymmetry, probes the distribution of linearly polarized gluons inside a transversely polarized proton, as well as the `Collins-like' fragmentation function, which is the analog of the Collins FF for gluon-originated jets.  In \cite{ref:500GeVCollins}, STAR provided the first limits on linearly polarized gluons in transversely polarized protons.  The Collins-like asymmetries presented here will provide far more restrictive upper limits.

The $\sin\left(\phi_S+\phi_H\right)$ and $\sin\left(\phi_S+2\phi_H\right)$ modulations of the cross section are sensitive to the transversity, Sivers and Boer-Mulders \cite{ref:BoerMuldersFunction} initial-state distributions convoluted with the Collins FF for the final-state. These, however, are expected to be negligible in all the kinematic configurations even when maximized scenarios for the distributions are considered \cite{ref:DAlesio2011}.  These terms were measured and found to be consistent with zero in the previous 500 GeV analysis \cite{ref:500GeVCollins}. In the 200 GeV $pp$ data reported here, they are also found to be consistent with zero with much smaller statistical uncertainties.

The remainder of this paper is arranged as follows: Section \ref{sec:Experiment} describes the subsystems of the STAR detector that are relevant for this measurement.  Section \ref{sec:DataSim} describes the data sets and simulation samples.  Section \ref{sec:JetsHadrons} describes the jet reconstruction and particle identification techniques used.  Section \ref{sec:Asymmetries} describes the single-spin asymmetry calculations, including the corrections that are applied to the data and the corresponding systematic uncertainties.  Section \ref{sec:Results} presents the results.  Finally, Sec.\@ \ref{sec:Conclusion} concludes.

\section{The STAR Detector at RHIC}\label{sec:Experiment}
RHIC at Brookhaven National Laboratory is the only accelerator in the world that is capable of colliding polarized proton beams, usually at center-of-mass energies of 200 or 510 GeV. It consists of two concentric, quasi-circular accelerator/storage rings on a horizontal plane. Each ring can store up to 120 bunches and is filled with 111 bunches under typical operations.

Each proton bunch may be given a different polarization direction, with a unique `spin pattern' assigned to groups of different bunches. These spin patterns can be varied each time the rings are filled. Over many fills, this leads to an equalization of spin states per bunch number, reducing spin-dependent systematic effects arising from bunch-to-bunch variations. 

The beam polarizations in RHIC are measured with proton-carbon (pC) Coulomb-nuclear interference polarimeters \cite{ref:pCPolarimeter} and a polarized atomic hydrogen jet (H-jet) polarimeter \cite{ref:HJetPolarimeter}. The pC polarimeters are fast detectors used to measure the relative polarization several times throughout a storage period or `fill', typically 8 hours long. The H-jet polarimeter provides an absolute measure of the beam polarization and is used to normalize the pC results. 

The Solenoidal Tracker at RHIC (STAR)~\cite{ref:StarOverviewNIM} is a multipurpose detector designed to measure the hadronic and electromagnetic particles emitted in heavy-ion and polarized proton-proton collisions. STAR comprises several subsystems that provide charged particle tracking and identification plus electromagnetic calorimetry over a wide range of pseudorapidity and full azimuth. The primary subsystems used for jet reconstruction in this work are the time projection chamber (TPC) \cite{ref:tpcNIM}, the barrel electromagnetic calorimeter (BEMC) \cite{ref:bemcNIM}, and the endcap electromagnetic calorimeter (EEMC) \cite{ref:eemcNIM}. Particle identification for the jet constituents is provided by the TPC and the barrel time-of-flight (TOF) detector \cite{ref:starTOF,ref:Llope:2012zz}.

The TPC provides charged particle tracking and identification in a 0.5~T solenoidal magnetic field for pseudorapidity $|\eta| \lesssim 1.3$ and full azimuthal angle ($\phi$). It determines the momentum of the outgoing charged particles and identifies the charged particles by measuring their ionization energy loss ($dE/dx$). The TPC is also used to locate the position of the collision vertex \cite{ref:tpcNIM}.

TOF measures the flight times of particles with a total timing resolution of $\sim80$ ps at $|\eta| \le 0.9$. It is constructed by a stack of resistive plates with five 220~$\mu$m gas gaps based on Multigap Resistive Plate Chambers (MRPCs) techniques, that were operated in a gas of 95\% R-134a and 5\% isobutane~\cite{ref:starTOF,ref:Llope:2012zz}. 


The BEMC is a lead-scintillator sampling calorimeter that surrounds the TPC in full azimuth over the range of $|\eta| < 1$. It is divided into 4800 towers of size $\Delta \eta \times \Delta \phi = 0.05 \times 0.05$, measures electromagnetic energy depositions and provides jet triggering in the experiment. The EEMC has a design similar to that of the BEMC, and extends the kinematic reach of the BEMC to $1.09 < \eta < 2$ with full azimuth. 

Two other global detector systems are important to this analysis. The vertex position detector (VPD) \cite{ref:vpdNIM} is a pair of timing detectors mounted directly around the beampipe that cover approximately half of the solid angle within the region $4.2 < |\eta| < 5.2$.  The VPD provides a minimally biased collision trigger and can be used to provide the start time for the TOF system.  The zero degree calorimeter (ZDC) \cite{ref:Adler:2000bd} is located in the region $|\eta| > 6.6$.  It is equipped with horizontal and vertical segmented scintillator shower maximum detectors (SMD), which are modeled after the EEMC SMD \cite{ref:eemcNIM}.  The ZDC SMD is used to verify the vertical orientation of the beam polarization at STAR.

\section{Data and Simulation}\label{sec:DataSim}
\subsection{Data sets and event selection}
The data used in this analysis were collected by the STAR experiment in 2012 and 2015 from transversely polarized $pp$ collisions at $\sqrt{s}$ = 200~GeV with integrated luminosities of 22 pb$^{-1}$ and 52 pb$^{-1}$, respectively.  The average polarization of the proton beams was about 57\% in both years, with fractional uncertainties of 3.5\% in 2012 and 3.0\% in 2015 \cite{ref:RHICPolG}.

The main physics triggers used to collect events for this analysis were jet patch (JP) triggers, which apply thresholds to the total transverse energy ($E_T$) observed within fixed $\Delta \eta \times \Delta \phi = 1 \times 1$ regions of the BEMC and EEMC.  There are 30 separate jet patches spanning the region $-1 < \eta < 2$, with five patches that overlap in $\eta$ for each of six non-overlapping regions in $\phi$. Two JP triggers were used during the 2015 running period, JP1 with a threshold of 5.4 GeV and JP2 with a threshold of 7.3 GeV.  Two additional triggers were utilized during the 2012 running period in order to provide better efficiency for low-$p_T$ jets, a JP0 trigger with a threshold of 3.5 GeV and a minimally biased collision trigger (VPDMB).  The VPDMB trigger required a coincidence between the VPD detectors at the two ends of STAR, with a timing cut to limit the location of the collision along the beamline. All JP2-triggered events were recorded. Non-JP2-triggered events that satisfied the other triggers were prescaled to fit within the available data-acquisition bandwidth.

Vertices were reconstructed from TPC tracks. If an event contains more than one candidate vertex, only the highest quality vertex, determined from the number of in-time TPC tracks and their transverse momenta, was considered.  For JP-triggered events, the $z$ position of the vertex along the beamline must be within 60 cm of the center of the TPC.  For VPDMB-triggered events, this range was reduced to 30 cm from the center of the TPC. Furthermore, for VPDMB-triggered events, the vertex $z$ positions measured by the TPC and VPD must satisfy $|z_{\text{tpc}} - z_{\text{vpd}}| < $ 6 cm. 

\subsection{Embedded simulation samples}\label{sec:embedding}
Monte Carlo simulations are needed to correct for detector effects on the measured quantities of interest, as well as to estimate various systematic uncertainties. Simulated events were generated using \textsc{Pythia} 6.4.28~\cite{ref:Pythia6} with the Perugia 2012 tune~\cite{ref:PerugiaTunes}. The parameter PARP(90), which controls the energy dependence of the underlying event process low-$p_{T}$ cut-off, was adjusted from 0.24 to 0.213 in Perugia 2012 in order to match previous STAR measurements of the $\pi^{\pm}$ cross sections in 200 GeV $pp$ collisions \cite{ref:starPion2005,ref:starPion2012}.  This modification was first introduced in Ref.\@ \cite{ref:STAR_2012_ALL}, where it was shown to provide a very good description of a wide range of jet and event properties in 510 GeV $pp$ collisions.  More recently, STAR has shown that this modified tune also provides a very good description of jet \cite{ref:STAR_groomed_jet,ref:STAR_invariant_mass} and underlying event \cite{ref:STAR_UE_paper} properties in 200 GeV $pp$ collisions. The simulated events were then processed through full detector simulations that match the detector configurations during 2012 and 2015 implemented in \textsc{Geant}~3~\cite{ref:Geant}. Event pile-up and beam background effects were incorporated by embedding the simulated events into real ‘zero-bias’ events collected during the 2012 and 2015 runs. The zero-bias events were triggered on bunch crossings over the span of the runs with a clock trigger, which was not correlated to the collisions.  The online trigger algorithm was replicated and applied to the embedding samples as well, allowing for an excellent reproduction of the spectra seen in the data.

The simulation software records the partonic hard scattering and the final-state particles from the fragmentation and hadronization of the partons, in addition to the response of the detector to those particles. These three distinct levels of information are referred to as the parton-level, particle-level, and detector-level, respectively. Parton- and particle-levels allow access to the full kinematics at the respective levels. The detector-level presents the simulated detector hit information in exactly the same format as is generated by real data events.

The configuration of the detector in 2012 is accurately implemented in \textsc{Geant}.  In contrast, the 2015 detector configuration in the \textsc{Geant} simulation does not fully represent the real 2015 STAR detector conditions. There are some minor differences due to the material, especially in the region $z \lesssim -30$ cm, where the support services for the Heavy Flavor Tracker (HFT)~\cite{ref:Bouchet_2009}, which was present during 2015, are not fully modeled. In order to account for this material difference, an additional simulation sample was produced using the 2012 detector configuration in the \textsc{Geant} simulation, but digitized using routines that describe the detector readout in 2015 and embedded into zero-bias events from 2015. Based on the comparison of the two simulation samples to data, the difference between these two configurations is believed to be substantially larger than the difference between the simulated 2015 detector configuration and the real detector. This additional sample was used for systematic uncertainty estimations in this analysis.

\section{Jet Reconstruction and Particle Identification}\label{sec:JetsHadrons}

The analysis of the 2012 data set largely follows the procedures outlined in Ref.~\cite{ref:500GeVCollins}. The higher statistics of the 2015 data set motivated a number of modifications in order to reduce systematic uncertainties.  The following sections focus on the procedures that are used in the 2015 data analysis, while noting those cases where the 2012 data analysis procedures differ significantly.

\subsection{Jet reconstruction}
The jet reconstruction procedures follow those of previous STAR jet analyses, and are used both in data and simulation. The anti-$k_{T}$ algorithm \cite{ref:AntiKtJets} implemented in the FastJet 3.0.6 package~\cite{ref:FastJet} with resolution parameter $R$ = 0.6 is used to reconstruct jets. Inputs to the jet finder include charged tracks from the TPC and calorimeter tower energies from the BEMC and EEMC. Tracks are required to have $p_{T} \ge 0.2$ GeV/$c$, and individual calorimeter towers need $E_{T} \ge 0.2$ GeV. Valid charged tracks are also required to contain more than twelve hits in the TPC and at least $51\%$ of the maximum number of points allowed by the TPC geometry and active electronic channels in order to provide good momentum resolution and remove split tracks. In order to remove pile-up tracks that are not associated with the hard scattering event, a $p_T$-dependent condition on the distance of closest approach (DCA) is imposed for tracks. Tracks with $p_T$ below 0.5 GeV/$c$ are required to have DCA $<$ 2~cm, while tracks with $p_T$ above 1.5 GeV/$c$ are required to have DCA $<$ 1~cm, with a linear interpolation in the intermediate $p_{T}$ region. To avoid double counting of the energy, all towers that have tracks pointing to them have the $p_{T}$ of the track (multiplied by $c$ to align the units) subtracted from the $E_{T}$ of the tower. If the track $p_{T} \cdot c$ is greater than the transverse energy of the tower, the tower $E_{T}$ is set to zero. This method has been shown to reduce the residual jet momentum corrections and the sensitivity to fluctuations in the hadronic energy deposition, resulting in an improved jet momentum resolution~\cite{ref:Adamczyk_2015_ALL}.

To be included in this analysis, jets are required to have a pseudorapidity $\eta$ (relative to the event vertex) between $-0.9$ and $0.9$, and a 'detector pseudorapidity' $\eta_{det}$ (relative to the center of STAR) between $-0.8$ and $0.9$. For the 2015 data analysis, jets containing tracks with $p_{T} > 20$ GeV/$c$ are rejected in order to reduce the sensitivity to decreasing track resolution at high momenta.  This cut is set at $p_T > 30$ GeV/$c$ for the 2012 data analysis. To suppress the possible contamination from non-collision background, such as cosmic rays and beam-gas interactions, the fraction of the jet energy in the calorimeters is required to be less than 0.95 and the transverse momentum sum of the charged tracks within a jet is required to be larger than $0.5$ GeV/$c$. Finally, jets reconstructed in jet-patch-triggered events are subject to additional constraints. To minimize trigger bias in the vicinity of the thresholds, every jet must match geometrically to a jet patch that could have triggered the event and exceed a minimum $p_{T}$ value of 6.0 GeV/$c$ (for JP0 and JP1) or 8.4 GeV/$c$ (for JP2).

\subsection{Hadron selection} \label{subsec:Hadron_selection}

For the measurements of hadron correlations in jets, hadrons are required to carry a fraction $z>0.1$ of the total jet momentum to reduce the contribution from underlying event particles.  The minimum distance between the hadron and jet directions in $\eta \times \phi$ space ($\Delta R_h$) is required to be larger than 0.05 to ensure valid $\phi_H$ reconstruction.  Note that this minimum $\Delta R_h$ requirement results in a cutoff at low hadron-$j_T$ values that rises as the hadron-$z$ and jet-$p_T$ increase. At least 50\% of the total TPC hits used in the track reconstruction are required to have valid $dE/dx$ information to provide good resolution for particle identification. The hadron momentum transverse to the jet axis ($j_{T}$) is also required to be smaller than 2.5 GeV/$c$ in addition to the upper limits from the underlying event studies (see Sec.\@ \ref{subsec:UE_corrs}).

Charged hadrons are only selected for the asymmetry analyses if the observed $dE/dx$ is consistent with the expected values for a particular particle. For example, pions are selected if the observed $dE/dx$ is consistent with the calculated value for pions of the measured momentum. A normalized $dE/dx$ is quantified as:
\begin{equation}
n_\sigma(\pi) = \frac{1}{\sigma_{exp}}\mathrm{ln}\left(\frac{dE/dx_{obs}}{dE/dx_{\pi,calc}}\right),
\end{equation}
where $dE/dx_{obs}$ is the observed energy loss of the tracks in the TPC, $dE/dx_{\pi,calc}$ is the expected energy loss for charged pions based on the Bichsel formalism \cite{ref:Bichsel:2006cs}, and $\sigma_{exp}$ is the $dE/dx$ resolution of the TPC~\cite{ref:PeakShift,ref:ShaoParticleId}. The value of $n_\sigma(\pi)$ is required to be within (-1, 2) in order to remove a large fraction of contamination from kaons, protons and electrons (see discussion in Sec.\@ \ref{sec:PID}). Similarly, non-overlapping $dE/dx$ regions can be chosen with the same kinematics (jet-$p_T$, hadron-$z$, hadron-$j_T$) where the observed $dE/dx$ is consistent with the expected values for charged kaons, protons and electrons, {\it e.g.}, $-5 < n_\sigma(\pi) < -1$, $2 < n_\sigma(\pi) < 4$ and $4 < n_\sigma(\pi) < 7$.

The $n_\sigma(\pi)$ for charged pions is expected to be a Gaussian distribution with a centroid value near zero and unit width. Similarly, $n_\sigma(K)$, $n_\sigma(p)$ and $n_\sigma(e)$ for kaons, protons, and electrons can be defined respectively. As with previously published STAR particle identification measurements~\cite{ref:PeakShift,ref:ShaoParticleId}, the determination of the centroid positions and widths of $n_\sigma(\pi,K,p,e)$ to high precision requires calibration. The pure samples of pions and protons are selected from topologically reconstructed $K_S$, $\Lambda$, and $\bar{\Lambda}$ decays to measure the  $dE/dx$ response over a wide momentum range, together with kaons identified by TOF over a narrower momentum range. The $dE/dx$ centroids and widths for all the hadrons follow universal curves as functions of $p/m = \beta\gamma$.  By selecting the tracks with $E/p$ close to unity, where $E$ is measured by the electromagnetic calorimeters and $p$ is measured by the TPC, highly enriched electron samples are obtained.  The resulting $n_{\sigma}(e)$ distributions are fitted with multi-Gaussian distributions, with fixed hadron centroids and widths, to obtain the centroid and width of the electron $dE/dx$ distributions as a functions of $p$.

\subsection{Underlying event corrections}\label{subsec:UE_corrs}

\begin{figure}[!hbt]
  \centering
    \includegraphics[width=1\columnwidth]{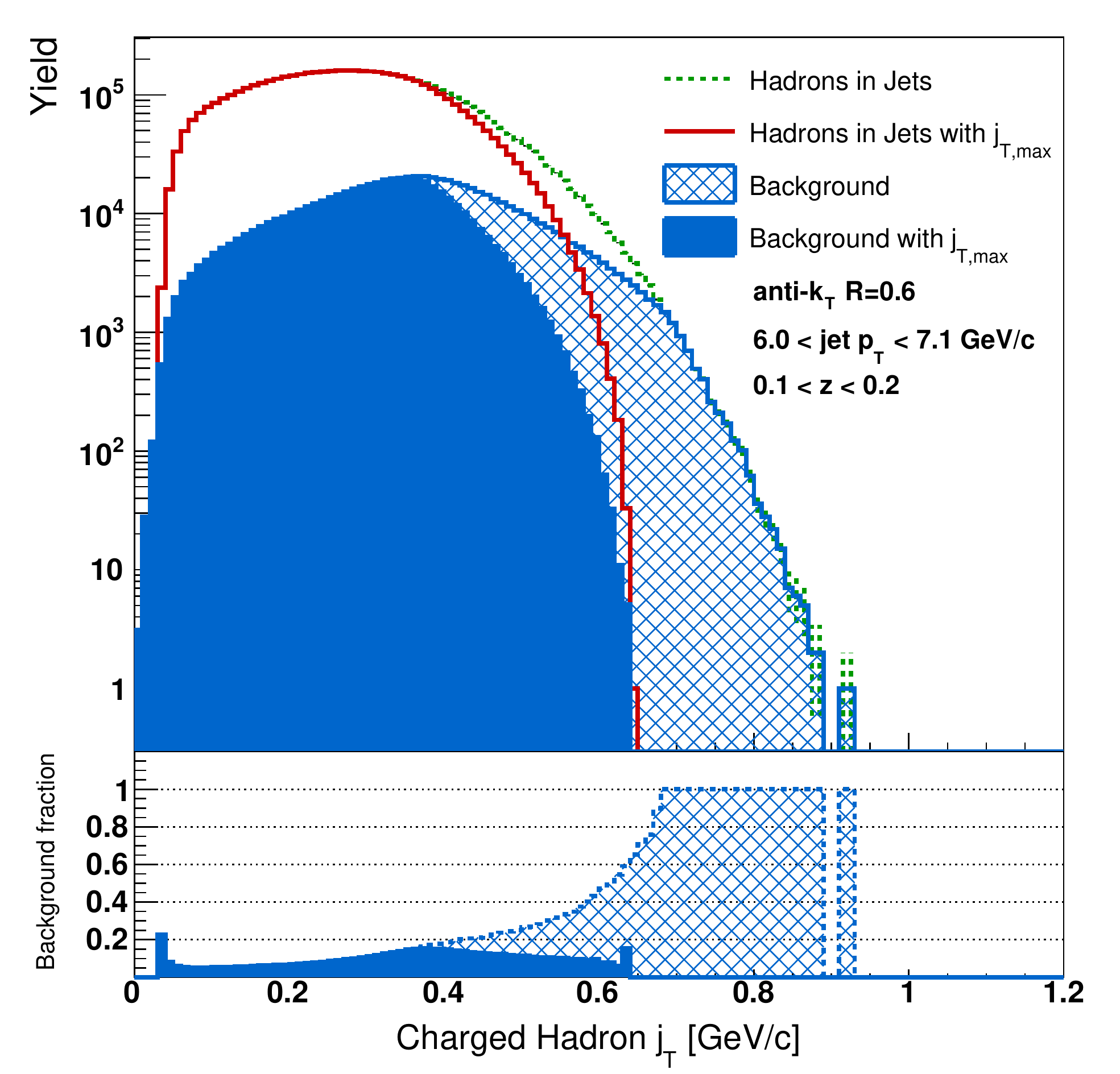}
    \caption{Distribution of the charged hadrons inside jets as a function of the hadron momentum transverse to the jet thrust axis, $j_{T}$, for jets with $6.0 < p_{T} < 7.1$ GeV$/c$ and hadrons with $0.1 < z < 0.2$. In the top panel, the lines show all the hadrons inside the jets, while the blue filled areas are the background from the off-axis cones before and after the $j_{T, max}$-cut described in Eq.~(\ref{equ:high_jT_cut}). The bottom panel shows the ratio of background over the hadrons inside the jets before and after the $j_{T,max}$-cut.}
    \label{fig:UE_jT_cuts_Example}
\end{figure}

Events with hard jets are often accompanied by a more diffuse background originating from multi-parton interactions or soft interactions between the scattered partons and beam remnants. These underlying event particles are unrelated to the hard partonic scattering of interest, but may contribute additional energy and transverse momentum to the reconstructed jets. At RHIC energies, the soft background particles of the underlying event are believed to be evenly distributed over $\eta$-$\phi$ space in $pp$ collisions~\cite{ref:STAR_UE_paper}, so the actual underlying event energy density is assumed to be uniform. The STAR detector has very good symmetry in $\phi$, but not in $\eta$, especially in the Endcap region where there is a service gap between the BEMC and EEMC. Because of these variations in detector performance with $\eta$, the off-axis cone method~\cite{ref:STAR_2012_ALL} is used to estimate the underlying event activity at the same $\eta$ as that of the jet under consideration, but at values of $\phi$ that are far from the hard jets in the event.\par

\begin{figure}[!hbt]
    \includegraphics[width=0.99\columnwidth]{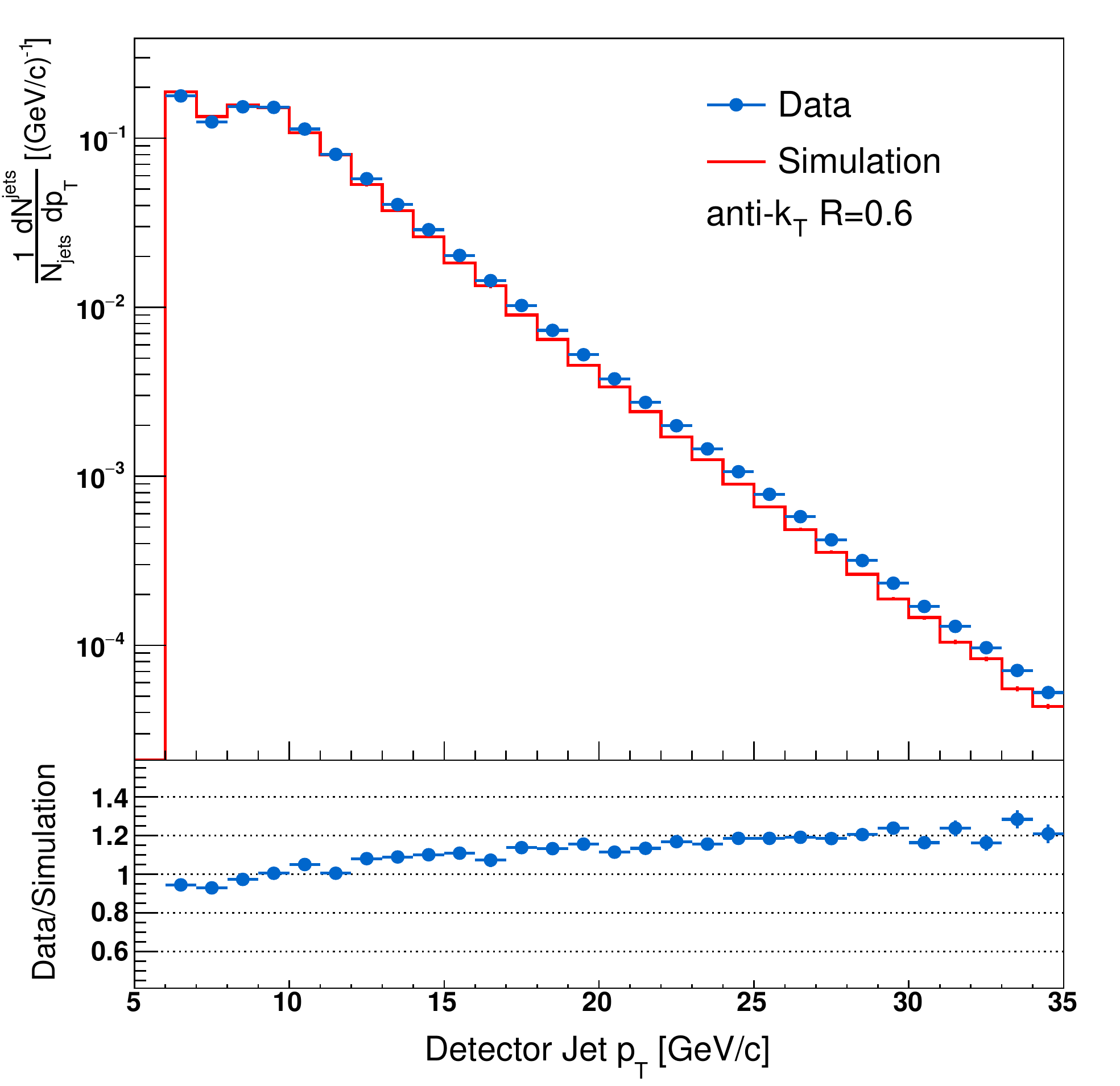}
    \caption{Distribution of the normalized jet yield as a function of detector jet-$p_{T}$ in 2015 data and simulation. The lower panel shows the ratio between data and simulation.}
    \label{fig:figure1}
\end{figure}

\begin{figure*}
\centering
\begin{minipage}{1\columnwidth}
  \centering
 \includegraphics[width=0.9\linewidth]{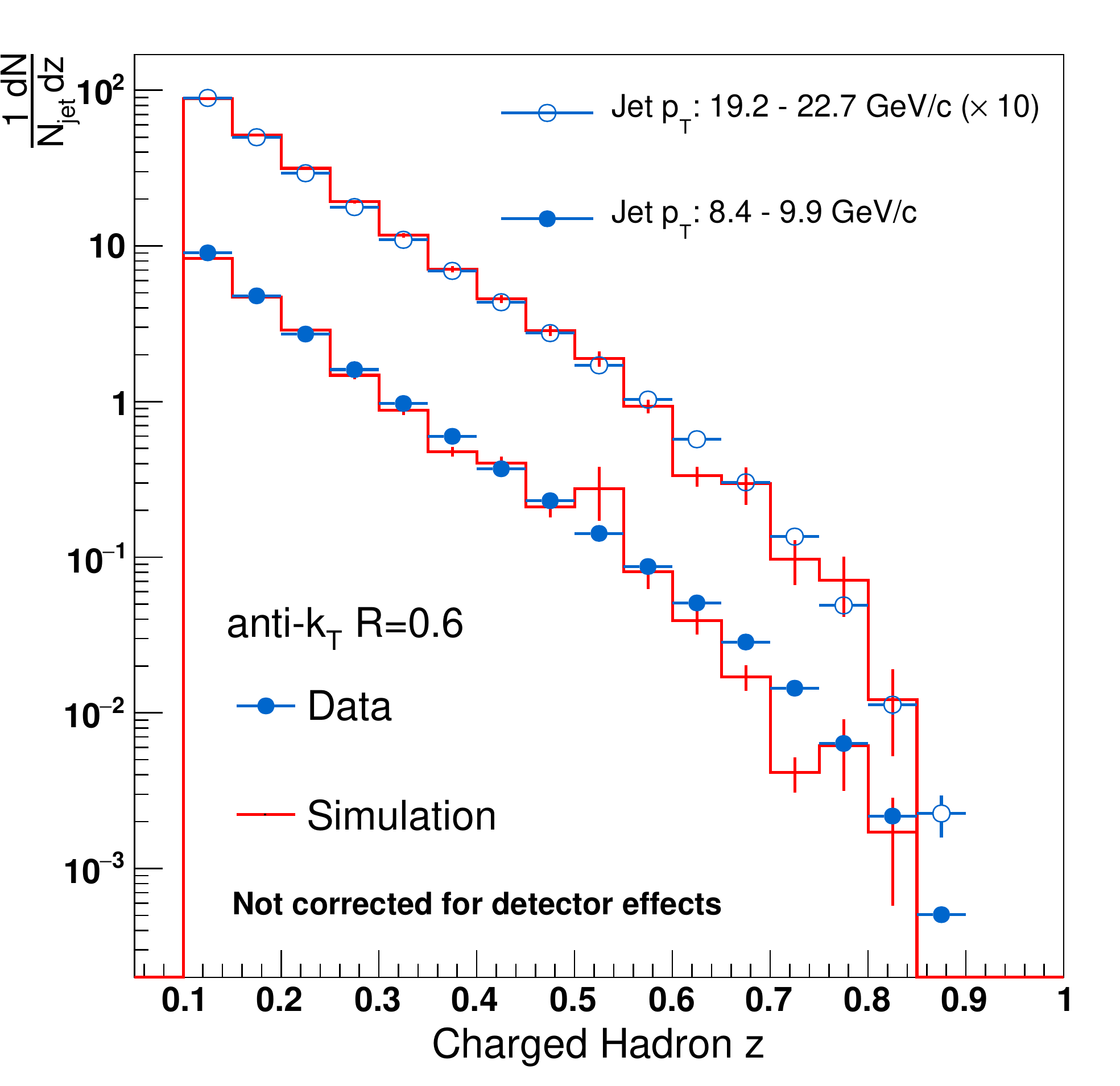}
  \caption{Comparison of data with simulation for charged hadrons within jets in the 2015 data as a function of the hadron longitudinal momentum fraction, $z$, in two different ranges of jet-$p_{T}$.}
  \label{fig:figure2_z}
\end{minipage}%
\hfill
\begin{minipage}{1\columnwidth}
  \centering
  \includegraphics[width=0.9\linewidth]{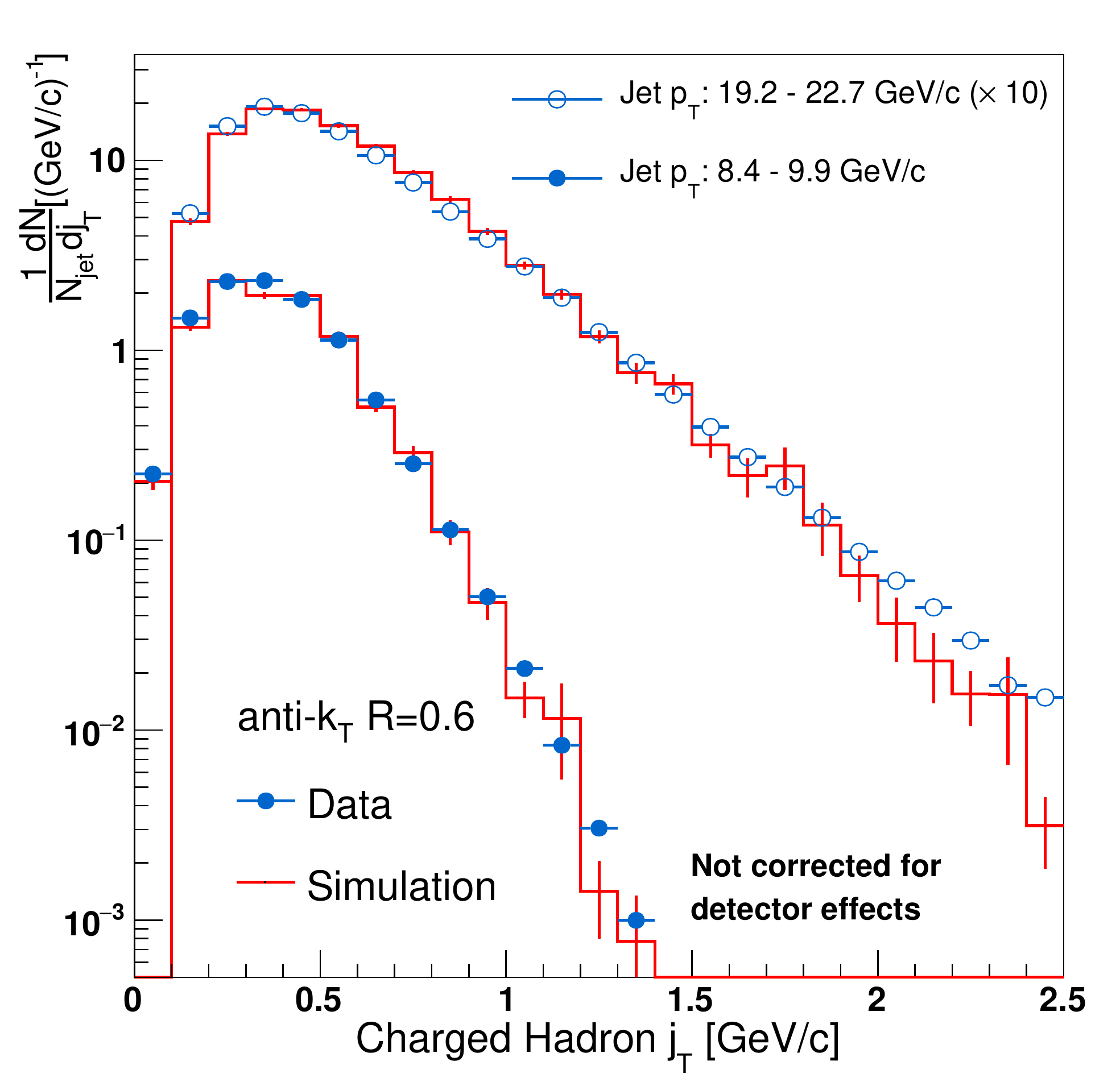}
  \caption{Comparison of data with simulation for charged hadrons within jets in the 2015 data as a function of the hadron momentum transverse to the jet axis, $j_{T}$, in two different ranges of jet-$p_{T}$.}
  \label{fig:figure2_jT}
\end{minipage}
\end{figure*}

The off-axis cone method is adapted from the perpendicular cones method used in the ALICE experiment~\cite{ref:ALICE_UE}, and has already been used in previous STAR jet and dijet measurements~\cite{ref:STAR_2009_Endcap_ALL,ref:STAR_2012_ALL,ref:STAR_2015results_ALL,ref:STAR_2013_ALL}.  In this method, two cones with radii equal to the jet resolution parameter ($R$ = 0.6) are defined for the reconstructed jet, with each cone centered at the same $\eta$ as the jet, but rotated $\pm\,\pi/2$ away in $\phi$. All particles falling within the two cones are collected, and the cone transverse momentum density ($\rho_{T,UE}$) is defined as the scalar sum of the $p_T$ of all the particles inside the two cones divided by their combined area, $2\pi R^2$. The jet transverse momentum $p_{T}$ is corrected with:
\begin{equation}
p_{T}^{corr} = p_{T} - \rho_{T,UE}\times A_{jet},
\end{equation}
where $A_{jet}$ is the area of the jet, which is calculated using the active area technique utilizing ghost particles~\cite{ref:Ghost_Particle} as implemented in the Fastjet package~\cite{ref:FastJet}. The ghost particles are thrown over a pseudorapidity range $|\eta| < 3$ and the ghost area is set to 0.04.\par

For the 2015 data analysis, the underlying event subtraction is applied to both data and simulation on the detector-level jet, and to simulation on the particle-level jet. For the 2012 data analysis, the underlying event correction is only applied to the particle-level jets. In this way, results from the two years are presented using the same jet momentum scale, even though intermediate steps in the two analyses are handled differently. Also, Ref.\@ \cite{ref:STAR_2012_ALL} found that particle jet-$p_T$ values calculated with underlying event subtraction are very close to those of their matching parton jets, which facilitates comparison to next-to-leading-order calculations.\par

The off-axis cone method is also used to study the underlying event activity inside the jets. The underlying event contamination is small in most kinematics regions; however, at large-$j_{T}$, low-$z$, and low jet-$p_T$, it can be significant. Figure~\ref{fig:UE_jT_cuts_Example} shows the distribution of charged hadrons inside jets as a function of $j_{T}$ for jets with $6.0 < p_T < 7.1$ GeV$/c$ and hadrons with $0.1 < z < 0.2$. The yields from the off-axis cones are normalized with $A_{jet}/(2\pi R^2)$. As can be seen, at $j_{T} >$ 0.6 GeV$/c$, almost all of the charged hadrons are from underlying events. Based on this observation, it is not practical to make Collins asymmetry measurements in this kinematic region. The $j_{T}$ distribution varies with hadron-$z$ and jet-$p_{T}$, so a $p_{T}$- and $z$-dependent $j_{T}$-cut is adopted in the Collins and Collins-like effect analyses. To limit the background inside the jets and minimize the uncertainty in underlying event subtraction, additional requirements are placed on the hadron-$j_{T}$:
\begin{equation}
\label{equ:high_jT_cut}
 \begin{split}
 j_{T} < & ~j_{T,max} = \\
 {  } & ~\mathrm{Min}[ (0.025+0.3295\times z)\times p_{T}^{jet}, ~2.5~\mathrm{GeV}/c],
 \end{split}
\end{equation}
The total background fraction in this kinematic range is reduced from 12.1\% to 8.5\%.  Furthermore, the underlying event background fraction is never larger than 20\% at any $j_T$ value. This upper $j_{T}$-cut has only a modest impact on the signal, but significantly reduces contributions from the underlying event.\par

It is important to note that Fig.\@ \ref{fig:UE_jT_cuts_Example} represents the worst case.  The mean $p_T$ of underlying event particles in 200 GeV $pp$ collisions is only 0.6 GeV/$c$ \cite{ref:STAR_UE_paper}. Thus, the underlying event fraction drops rapidly as jet-$p_T$ and/or hadron-$z$ increase.  It drops to less than \,2\% for $p_T > 11.7$ GeV/$c$ and $0.1 < z < 0.2$. It is under \,2\% at all $p_T$ for $z > 0.2$.

\subsection{Comparison to simulation}\label{sec:Compare_to_sim}

In the simulation samples, detector-level jets are reconstructed from the simulated TPC and calorimeter responses using the same algorithms as real data. Figure~\ref{fig:figure1} compares the distribution of the normalized jet yields as a function of detector jet-$p_{T}$ in data and simulation for 2015. Figures~\ref{fig:figure2_z} and \ref{fig:figure2_jT} show the distributions of charged hadrons within jets in the 2015 data as a function of the charged hadron longitudinal momentum fraction $z$ and charged hadron momentum transverse to the jet axis, $j_{T}$, in two representative detector jet-$p_{T}$ ranges, 8.4 - 9.9 GeV/$c$ and 19.2 - 22.7 GeV/$c$. There is good agreement between data and simulation for these quantities.  Similar agreement is also seen in the 2012 data.  These comparisons indicate that the detector conditions are well reproduced in the simulation.\par

Jets are also reconstructed in simulation at the particle- and parton-level using the anti-$k_{T}$ algorithm with $R = 0.6$. Particle-level jets are reconstructed from all final-state stable particles including those arising from the underlying event and beam remnants. Parton-level jets are reconstructed from the hard-scattered partons including those from initial- and final-state QED and QCD radiation, but not those from beam remnants or underlying event effects.\par

The detector-level jets are influenced by finite resolutions and efficiencies of the detector. Thus, to estimate corrections and systematic uncertainties, it is important to correlate the jets reconstructed at the particle- or parton-level to the simulated jets reconstructed at the detector-level. In this way, the original jet properties are correlated to the ones reconstructed in the detector. For hadrons within jets, the correlation of individual tracks to the particles inside the matched particle-level jet is also useful in this analysis. In practice, for jet matching, a particle- or parton-level jet is associated with a jet at the detector-level if the distance between the jet axes is within $\Delta R = \sqrt{\Delta\eta^2 + \Delta\phi^2} < 0.5$. The closest parton or particle jet is chosen if more than one matched jet is found. Hadrons are matched between jets by finding the closest tracks with $\Delta R < 0.04$ and the same charge.\par

\section{Transverse Single-Spin Asymmetries}\label{sec:Asymmetries}

Asymmetries are extracted using the `cross-ratio' formalism \cite{ref:OhlsenCR} to cancel detector efficiencies to leading order and eliminate the need for spin-dependent luminosity factors:
\begin{widetext}
\small
\begin{equation}
    \label{AN}
    A_{N}\sin(\phi) = \frac{1}{P}\cdot\frac{\sqrt{N^{\uparrow}(\phi)N^{\downarrow}(\phi+\pi)} - \sqrt{N^{\downarrow}(\phi)N^{\uparrow}(\phi+\pi)}}{\sqrt{N^{\uparrow}(\phi)N^{\downarrow}(\phi+\pi)} + \sqrt{N^{\downarrow}(\phi)N^{\uparrow}(\phi+\pi)}} = \frac{\sqrt{\sum_{i}P_i N_{1,i} \cdot \sum_{i}P_i N_{2,i}} - \sqrt{\sum_{i}P_i N_{3,i} \cdot \sum_{i}P_i N_{4,i}}}{\sqrt{\sum_{i}P_i^{2} N_{1,i} \cdot \sum_{i}P_i^{2} N_{2,i}} + \sqrt{\sum_{i}P_i^{2} N_{3,i} \cdot \sum_{i}P_i^{2} N_{4,i}}},
\end{equation}
\end{widetext}
where $N^{\uparrow}$ (or $N^{\downarrow}$) is the yield for a given spin state in each detector half weighted by the beam polarization ($P$). To account for slightly varying conditions and the slow polarization decay over time~\cite{ref:RHICPolG}, events are weighted by the beam polarization per run ($i$) for each spin state and azimuthal region ($N_1$...$N_4$).

The asymmetry modulations are independently extracted by binning Eq.~(\ref{AN}) in  the appropriate $\phi_{S}$ and $\phi_{H}$ combination and fitting the result with the sinusoidal function:
\begin{equation}
    \label{equ:asym_fit}
    p_{0} + p_{1}\times \sin(\phi),
\end{equation}
where $p_{1}$ is the asymmetry and $p_{0}$ is an offset.  For the $A_{UT}^{\sin(\phi_S)}$ measurements, six $\phi_S$ bins spanning the range ($-\pi/2$, $\pi/2$) are used, while for the Collins and Collins-like asymmetry measurements, twelve $\phi_C = \phi_S-\phi_H$ and $\phi_{CL} = \phi_S - 2\phi_H$ bins spanning ($-\pi$, $\pi$) are used. In all three cases, the $\chi^{2}$ distributions of the fits are consistent with the expected distributions with the same number of degrees of freedom. Both beams are polarized during the collision, so each jet is analyzed twice in the asymmetry calculation. Depending on the beam direction and the pseudorapidity of the jets, the asymmetries are calculated both for jets scattered forward ($x_{F} > 0$) and backward ($x_{F} < 0$) relative to the polarized beam. The yields from both beams are merged to maximize the statistical precision.\par

\subsection{Underlying event dilutions}
As has been discussed in Sec.\ IV, for certain kinematics, the underlying event contamination can be as large as 20\%, so it is not negligible in the asymmetry calculation. $A_{UT}$ was also calculated for hadrons in the off-axis cones and found to be consistent with zero.  Thus, in the 2015 data analysis the underlying event contributions are treated as simple dilutions to the measured asymmetry values. In each kinematic region with the underlying event fraction $f_{UE}$, the measured asymmetry is corrected by:
\begin{equation}
A_{N} \rightarrow A_{N,corr} = \frac{A_{N}}{1-f_{UE}},
\end{equation}
where $A_{N}$ is the measured asymmetry, and $A_{N,corr}$ is the asymmetry after the underlying event correction. The statistical uncertainty on $f_{UE}$ is negligible compared to its systematic uncertainty. The underlying event yields were also evaluated by comparing the background fractions for jets reconstructed using the anti-$k_{T}$ and $k_{T}$ algorithms. The anti-$k_{T}$ algorithm is less likely to cluster soft underlying event particles during jet reconstruction, and its jet shapes are more regular than those of the $k_{T}$ algorithm. The $f_{UE}$ values for the two algorithms differ by up to 10\%, depending on the kinematics. This difference is added as an additional systematic uncertainty to the measured asymmetries, which is less than 10\% of the statistical uncertainty in all cases.\par

Given the lower statistical precision of the 2012 data, underlying event corrections were not applied at the detector-level, as noted in Sec.\@ \ref{subsec:UE_corrs}.  Rather, a systematic uncertainty is included to account for the underlying event dilution of the measured asymmetries. Typically, the systematic uncertainties are less than 30\% of the statistical uncertainty, except for low-$p_T$, low-$z$, and high-$j_T$, where it can reach up to 60\% of the statistical uncertainty.

\subsection{Particle contamination}\label{sec:PID}

\begin{figure}[!hbt]
    \includegraphics[width=1.0\columnwidth]{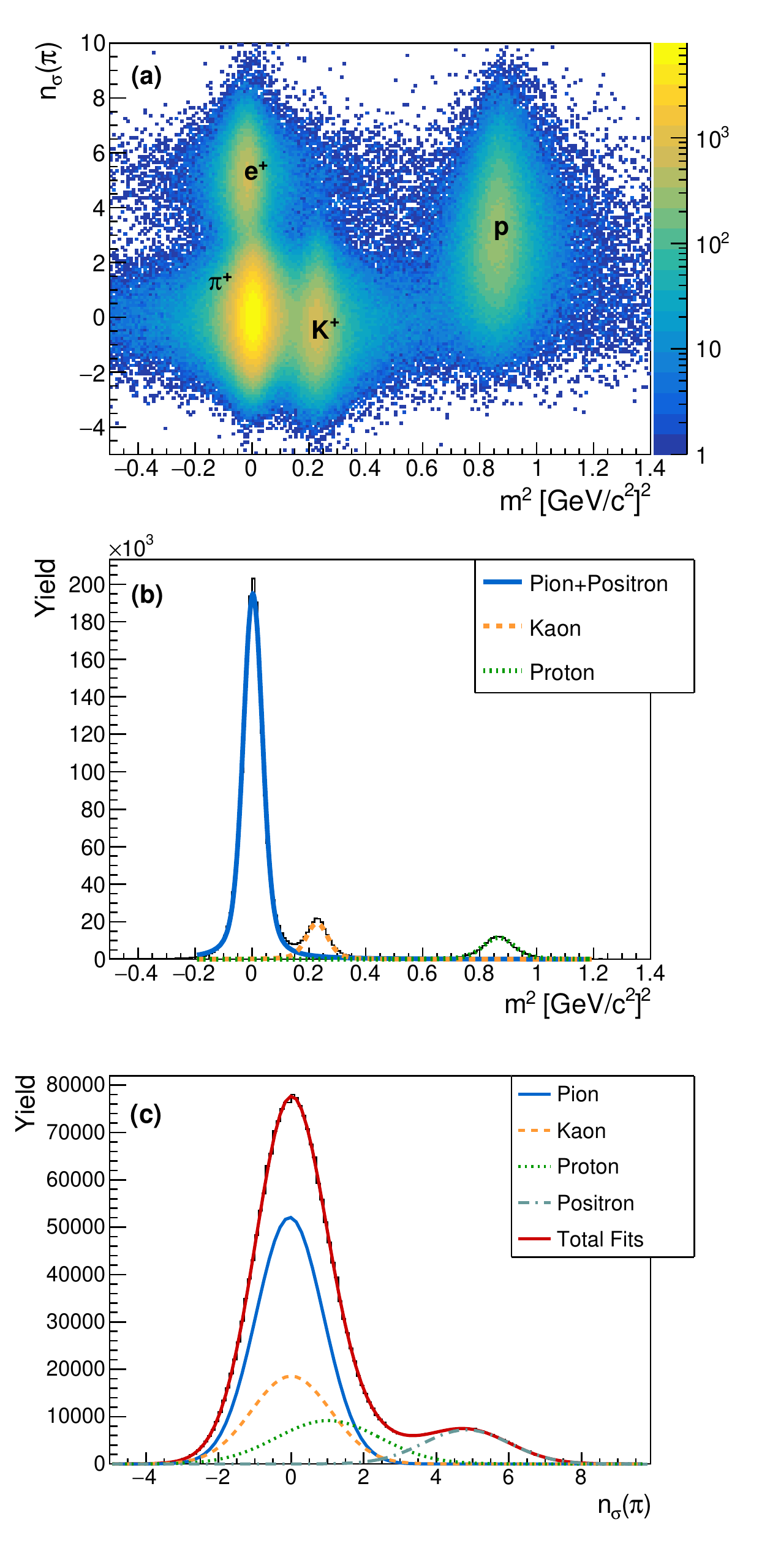}
    \caption{(a): The correlations of $n_\sigma(\pi)$ \textit{vs}.\@ $m^{2}$ for positively charged particles carrying momentum fractions of 0.1 $< z <$ 0.13 in jets with 8.4 $< p_{T} <$ 9.9 GeV$/c$. (b): Projection to the $m^{2}$ distribution with Multi-Voigt profile fits. (c): Projection to the $n_\sigma(\pi)$ distribution with Multi-Gaussian fits.}
    \label{fig:dEdx_tofm2_2D}
\end{figure}

The hadron-in-jet asymmetries are presented separately for identified charged pions ($\pi$), kaons ($K$) and protons ($p$). In each case, a $\pi/K/p$ rich sample is first identified using the $dE/dx$ information from the TPC discussed in Sec.~\ref{subsec:Hadron_selection}. The raw asymmetries ($A_N$) extracted from these enriched samples are linear mixtures of the pure $\pi/K/p$ asymmetries ($A_{N,pure}$). Note that all of the enriched samples are also contaminated by electrons produced by photon conversions and semi-leptonic heavy flavor decays. The photons arise primarily from $\pi^{0}$ decay.  The $\pi^0$ asymmetries are expected to be nearly zero as they are close to the average of the $\pi^{+}$ and $\pi^{-}$ asymmetries. Heavy flavor production arises primarily from the $gg \rightarrow q\bar{q}$ process, which does not contribute to either the Collins or Collins-like effect. Indeed, electron asymmetries measured using samples integrated over wide ranges of jet-$p_{T}$, hadron-$z$ and $j_{T}$, are found to be consistent with zero within the statistical uncertainties. For these reasons, the electron asymmetries are assumed to be zero in the following discussion.\par

The relationship between the raw and pure asymmetries is given by $A_{N} = A_{N,pure} M$, and therefore $A_{N,pure} = A_{N} M^{-1}$, where $A_N, A_{N,pure}$ and $M$ are defined as:

\begin{equation}
\label{Equ:A_raw}
A_{N} = (A_{\pi_{rich}}, A_{K_{rich}}, A_{p_{rich}}),
\end{equation}
\begin{equation}
\label{Equ:A_pure}
A_{N,pure} = (A_{\pi}, A_{K}, A_{p}),
\end{equation}
\begin{equation}
\label{Equ:Matrix_frac}
  M = \\
  \begin{pmatrix}
    f_{\pi_{rich}}^{\pi} & f_{K_{rich}}^{\pi} & f_{p_{rich}}^{\pi} \\
    f_{\pi_{rich}}^{K}   & f_{K_{rich}}^{K}   & f_{p_{rich}}^{K}   \\
    f_{\pi_{rich}}^{p}   & f_{K_{rich}}^{p}   & f_{p_{rich}}^{p}   \\
  \end{pmatrix},
\end{equation}

\noindent here, $f_{i_{rich}}^{j}$ is the fraction of particle type $j$ in the $i$-rich sample, with $f_{i_{rich}}^{\pi} + f_{i_{rich}}^{K} + f_{i_{rich}}^{p} = 1 - f_{i_{rich}}^{e}$. The particle fractions are calculated using both the TOF and TPC information. As with the selection of the hadron-rich samples, the $dE/dx$ from the TPC is used for $\pi, K, p$ and $e$ identification over the entire momentum range. However, due to the characteristic variation of $dE/dx$ with particle momentum, the $dE/dx$ of two different particle types will overlap in some kinematic regions. For example, at particle momentum $\sim$\,1.1 GeV$/c$, kaons and pions have the same $dE/dx$; while at momentum $\sim$\,1.7 GeV$/c$, protons and pions experience the same $dE/dx$; kaons and protons have the same $dE/dx$ at $\sim$\,2.5 GeV$/c$. In these regions, the TOF can provide additional discrimination power for particle identification. \par

The TOF measures the arrival time of the particles.  TPC tracks provide both the path length from the collision vertex to the TOF hit and momentum measurement for the associated particles. With the time of the collision measured by the VPD ($t_{start}$), the inverse velocity $1/\beta$ and then the mass-squared $m^2$ are calculated. The VPD consists of identical scintillation detector arrays located close to the beam pipe at each end of the STAR detector. It measures the collision time in 200 GeV $pp$ collisions with a resolution of $\sim$\,80 ps by detecting the forward high energy particles from the collisions. It was found that the VPD efficiency for getting the correct start time was low at the high instantaneous luminosities experienced during the 2015 data taking period. Only $\sim$\,10\% of the TPC tracks had valid TOF information when using the VPD to measure $t_{start}$. In order to improve the efficiency, a method called the `start-less TOF' algorithm was adopted for the 2015 data analysis. Previously, STAR had only used start-less TOF when analyzing data from low-energy Au+Au collisions. The approach assumes that any track with 0.2 $< p <$ 0.6 GeV/$c$ and within two standard deviations of the expected pion $dE/dx$ value is a pion. Then the start time of the event is taken to be the average start time of these pions based on their mass, momenta, and path lengths.\par

Figure~\ref{fig:dEdx_tofm2_2D} shows the correlation between $n_\sigma(\pi)$ and $m^{2}$ in 2015 data for tracks carrying longitudinal momentum fraction 0.1 $< z <$ 0.13 in jets with 8.4 $< p_{T} <$ 9.9 GeV$/c$. There is a very good separation of kaon and pion $m^{2}$ values in the region where they overlap in $n_\sigma(\pi)$. The surrounding background is suppressed by two to three orders of magnitude relative to the mass peak, and the fraction of tracks with valid TOF information is increased significantly compared to the measurement using the VPD. With these improvements, pions and kaons can be identified with TOF up to $p_{T} \sim 1.6$ GeV/$c$, and protons can be separated from $\pi + K$ up to $p_{T} \sim 3.0$ GeV/$c$~\cite{ref:PeakShift}.\par

\begin{figure}[!hbt]
    \includegraphics[width=1.0\columnwidth]{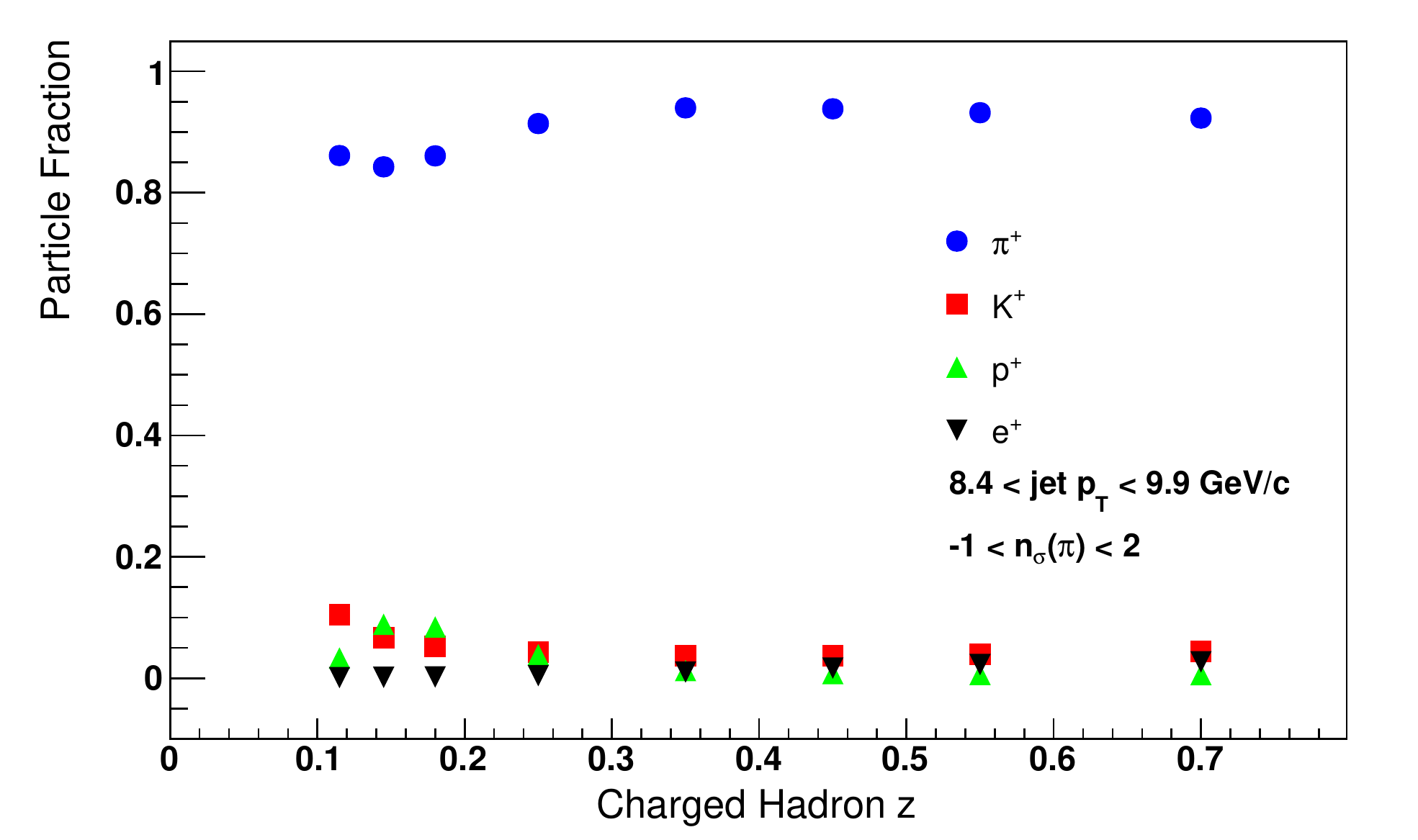}
    \caption{Charged particle fractions as a function of the hadron longitudinal momentum fraction, $z$, for charged particles that satisfy $-1 < n_\sigma(\pi) < 2$ (pion-rich region), in jets with 8.4 $< p_T <$ 9.9 GeV$/c$. The error bars are within the points.}
    \label{fig:figure_pikpe_frac}
\end{figure}

The particle fractions in each jet-$p_{T}$ and hadron-$z$ kinematic bin are extracted by taking ratios of the $\pi/K/p/e$ yields determined by fitting the $m^{2}$ distribution with a multi-Voigt profile and the $n_{\sigma}(\pi)$ distribution with a multi-Gaussian function. This generally follows similar procedures as the previously published results~\cite{ref:500GeVCollins} but with improved precision. The TOF $m^{2}$ distributions are fit with a sum of three Voigt profiles, one each for pions+electrons, kaons, and protons as shown in the middle panel of Fig.~\ref{fig:dEdx_tofm2_2D}. The electrons that fall under the pion peak in $m^2$ are well-separated through the $n_{\sigma}(\pi)$ cut ({\it e.g.}, $n_{\sigma}(\pi) < 3$ in Fig.~\ref{fig:dEdx_tofm2_2D}). The TPC $n_{\sigma}(\pi)$ distributions are fit with the sum of four Gaussian distributions, representing yields of pions, kaons, protons, and electrons as shown in the bottom panel of Fig.~\ref{fig:dEdx_tofm2_2D}, where the centroids and widths of the Gaussian distributions are fixed at the values determined by the calibration procedure discussed in Sec.~\ref{subsec:Hadron_selection}. Figure~\ref{fig:figure_pikpe_frac} shows the particle fractions in the pion-rich regions as a function of $z$ for jets with 8.4 $< p_{T} <$ 9.9 GeV$/c$. The uncertainties in the particle fraction measurements are dominated by systematic effects associated with fluctuations of the fitting parameters. A variation of the fitting parameters is considered, and the differences of the asymmetries measured with this variation are assigned as the final systematic uncertainties on particle identification.\par

In some kinematic bins, {\it e.g.}, in Fig.~\ref{fig:dEdx_tofm2_2D}, different particle-rich samples cannot be separated using $dE/dx$ alone, so only a single merged sample is defined, {\it i.e.}, $-5 < n_\sigma(\pi) < 2$ as a pion+kaon rich sample. The method for the pure asymmetry calculation is modified in these situations. As shown in Fig.~\ref{fig:dEdx_tofm2_2D}, when pions and kaons are located at similar values of $n_\sigma(\pi)$, they are well separated by TOF. So a pure kaon sample can be selected with $0.15 < m^{2} < 0.35\ \mathrm{GeV}^{2}/c^{4}$ and $-2.5 < n_\sigma(\pi) < 2.5$. This selected sample is defined as the pure kaon sample, and the calculated asymmetry is the pure kaon asymmetry. This means that $A_{K_{rich}} = A_{K}$, and the second matrix column in Eq.~(\ref{Equ:Matrix_frac}) becomes (0,1,0).  The events from the pure kaon sample are excluded from the other three particle-rich samples, then the rest of the analysis proceeds normally.  For cases where pions and protons or kaons and protons have the similar values of $dE/dx$, TOF is used to identify pure proton samples in the same manner.

The 2012 data analysis differed from the above in two ways.  First, the instantaneous luminosities during the 2012 data taking period were lower, so the VPD was used to provide start times for the TOF measurements.  Second, the reduced statistics meant that no attempt was made to separate kaons and protons.  Therefore, only the 2015 data are used to determine the Collins asymmetries for kaons and protons in Sec.\@ \ref{subsubsec:K_p_Collins}. For the high-$z$ charged pion inside jet $A_{UT}^{\sin(\phi_S)}$ measurement (Sec.\@ \ref{subsec:inclusive_asym}), the $dE/dx$ requirement was sufficient for enhancing the $u$ (for $\pi^{+}$) and $d$ (for $\pi^{-}$) quark fractions in the jet production.\par

\subsection{Kinematic corrections}
\label{subsec:kinematic_correction}
\subsubsection{Jet $\eta$ correction}

\begin{figure}[!hbt]
    \includegraphics[width=1.0\columnwidth]{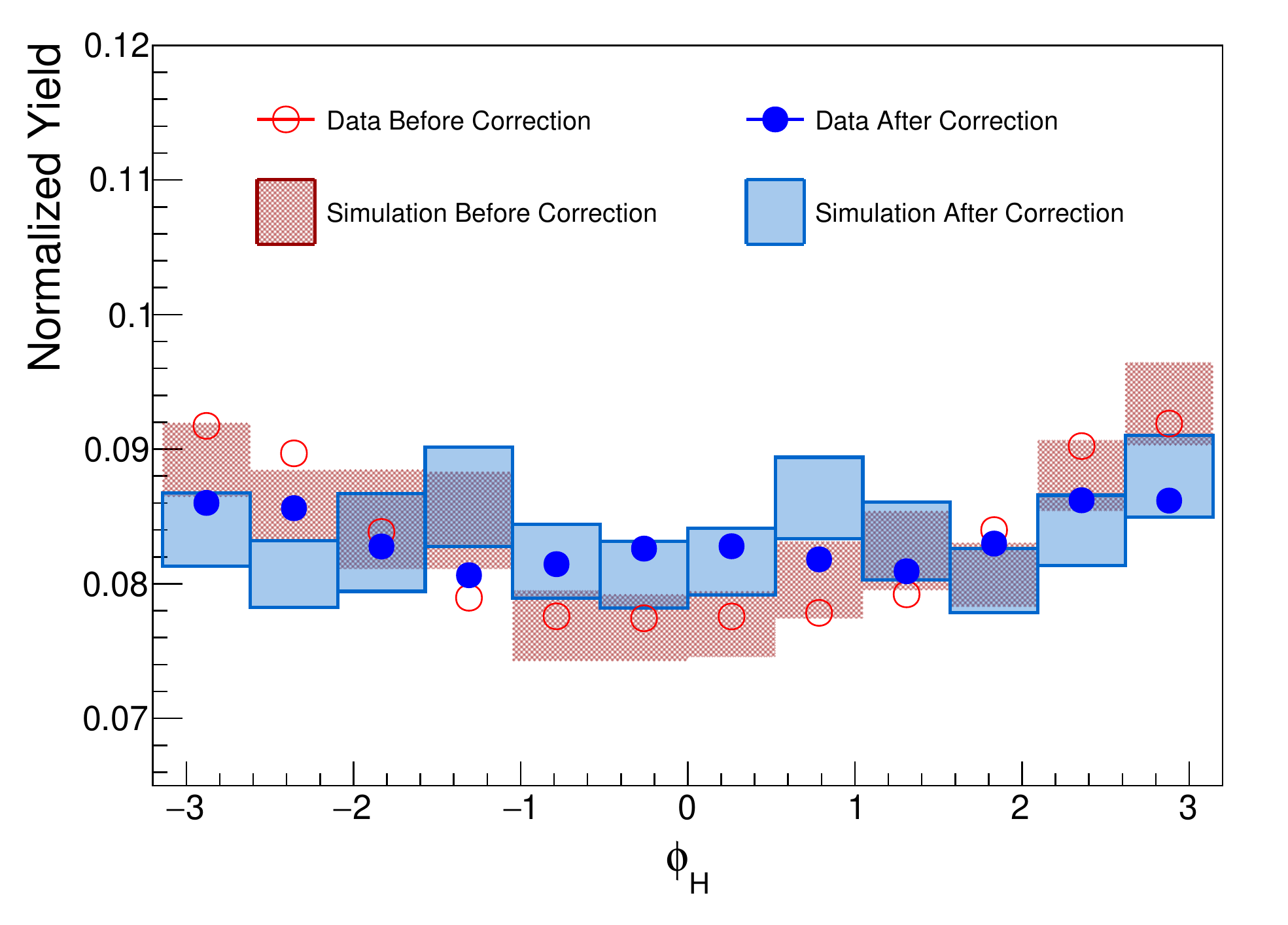}
    \caption{Distribution of the normalized charged hadron yields as a function of the azimuthal angle of the charged hadrons relative to jet scattering plane, $\phi_{H}$, in 2015 data and simulation. The data uncertainties are smaller than the points, while the boxes indicate the size of the simulation uncertainties.}
    \label{fig:figure_PhiH_EtaCorr}
\end{figure}

A study of nonuniform acceptance effects (discussed in Sec.\@ \ref{sec:leak_through}) in the 2015 data revealed that the observed large Collins asymmetries introduced a significant distortion to the Collins-like asymmetry measurements (up to 30\% of their statistical uncertainties), while all other intermodulation cross-talk effects were negligible. Upon detailed investigation, a small deviation in the reconstruction of the jet pseudorapidity near the detector edges was observed.  The small, but systematically consistent missing energy at larger pseudorapidity biased the reconstruction of the jet axis toward midrapidity.  This in turn biased the reconstruction of $\phi_H$. Figure~\ref{fig:figure_PhiH_EtaCorr} shows the distribution of the normalized charged hadron yields as a function of $\phi_{H}$ in both data and simulation. As can be seen, the reconstructed hadron was shifted towards the midrapidity side of the jet. 

The difference of the physics $\eta$ between the detector-level jet and its matched particle-level jet, $\delta \eta = \eta^{par} - \eta^{det}$, is calculated in the simulation. For each reconstructed detector-level jet from data or simulation, the pseudorapidity is corrected by $\eta + \delta \eta$. This correction is jet-$p_{T}$ and detector-$\eta$ dependent. The closed points in Fig.~\ref{fig:figure_PhiH_EtaCorr} show the same distribution of $\phi_{H}$ in both data and simulation after the $\eta$ correction. $\phi_{H}$ is more evenly distributed after the correction as the instrumental non-uniformity is reduced. This correction reduces the cross-talk from the Collins effect to the Collins-like effect in the 2015 data by a factor of three, corresponding to $<$\,2\% cross-talk.\par

The instantaneous luminosities were lower during the 2012 running period than during 2015, and without the HFT there was much less material inside the STAR detector.  Therefore, the tracking efficiency is higher in the 2012 data and falls off more slowly with pseudorapidity near the edges of the TPC. This, in addition to the reduced statistical accuracy in 2012 compared to 2015, means that the bias is not significant. Thus, the jet $\eta$ correction is not applied to the 2012 data.

\subsubsection{Correction to jet and hadron kinematics}
The transverse single-spin asymmetries vary slowly and approximately linearly over the full kinematic range, thus bin-by-bin unfolding is sufficient to correct the measured jet-$p_{T}$ or charged hadron-$z$ and $j_{T}$ for detector resolution and efficiency effects.\par

In order to compare the experimental results with theoretical predictions, a correction to the particle-level kinematics is made by applying kinematic shifts to the data points. The kinematic shifts are calculated with the same procedure as in the 2011 data analysis \cite{ref:500GeVCollins}.  Jets in the simulation and hadrons within those jets are matched between the detector- and particle-levels using the procedure described in Sec.\@ \ref{sec:Compare_to_sim}.  The average differences in the jet-$p_T$ values at the particle- \textit{vs}.\@ detector-levels, $\langle \delta p_T \rangle= \langle p_T^{par} - p_T^{det} \rangle$, are added to the measured jet-$p_T$ values in the data to correct them to the particle-level.  The statistical uncertainty from the embedding is treated as one of the systematic uncertainties on the jet-$p_T$.  The same procedure is used to correct the measured hadron-$z$ and $j_T$ values to the particle-level.

\subsubsection{Systematic uncertainties in the jet and hadron kinematics}
In order to estimate the systematic uncertainty associated with the detector efficiency, within the simulation, 4\% of tracks are randomly rejected before reconstructing jets. The difference in $\delta p_{T}$ between this sample and the nominal one is assigned as a systematic uncertainty on the kinematic shift. Additionally, for the 2015 data, $\delta p_{T}$ is also calculated using the simulation generated with 2012 detector configurations for the 2015 analysis (as discussed in Sec.\@ \ref{sec:embedding}). The difference in $\delta p_{T}$ between the two simulation samples is added as a systematic uncertainty on the kinematic shift.

Additional systematic uncertainties on the jet and hadron energy scales consist of two parts: one from the gain calibration and status uncertainties of the EMC towers, and the other from the TPC track transverse-momentum uncertainty and the uncertainty in the tower response to charged hadrons. The EMC gain calibration uncertainty is estimated to be 3.4\% in 2015 (3.8\% in 2012), and in both datasets the status uncertainty is estimated to be 1\% based on how well the monitoring software kept up with the failed channels. The track momentum uncertainty is found to be very small (0.3\%) from weak decay studies. The EMC tower hadron response uncertainty is taken as 1.4\%. These uncertainties are rescaled by the observed electromagnetic and hadronic energy fractions in the jets, then added in quadrature to the $\delta p_T$ uncertainties described above to obtain the total systematic uncertainties in jet-$p_T$.  The jet-$p_T$ uncertainty is then propagated as an additional contribution to the uncertainty in the hadron-$z$ determinations.

\subsection{Trigger bias}
In proton-proton collisions, quark-quark ($qq$), quark-gluon ($qg$) and gluon-gluon ($gg$) are the three dominant partonic scattering processes. The STAR jet patch trigger system may be more efficient for certain processes, which will alter the subprocess fractions and thus distort the measured asymmetries. In the simulation, detector-level jets are matched to particle- and parton-level jets. Then the parton-level jets are further matched back to hard-scattered partons from \textsc{Pythia} using the same $\Delta R$ cut to sort the events into quark, anti-quark, and gluon jets based on the \textsc{Pythia} record. In this way, the quark and gluon fractions observed at the detector-level can be calculated. The same procedure is used to match the particle-level jets back to hard-scattered partons in an unbiased pure \textsc{Pythia} sample. The pure \textsc{Pythia} sample is generated before adding the \textsc{Geant} model and trigger filter, and the final outputs are particle-level jets. The selection cuts on jet-$p_{T}$ and physics $\eta$ are retained for the pure \textsc{Pythia} study. $\Delta R_h$ between the single particle and parent jet axis is still required to be larger than 0.05, and the upper $j_{T}$-cut is also kept in the analysis.  All other detector related cuts are omitted.\par

The ratios of the biased-to-unbiased quark and gluon jet fractions are calculated, then the systematic uncertainties are evaluated as:
\begin{equation}
    \sigma_{Trig. Bias} = (|1-\mathrm{ratio}|) \times \mathrm{Max}(|A|,\sigma_A),
\end{equation}
where $A$ is the measured asymmetry and $\sigma_A$ is its statistical uncertainty.
The inclusive jet and Collins-like asymmetries are sensitive to gluons, while the Collins and jet with high-$z$ charged pion asymmetries are sensitive to quarks, so the trigger bias is calculated separately for the different processes. The trigger bias is the dominant systematic uncertainty in this analysis, and the ratio of the biased-to-unbiased quark and gluon jet fractions can vary up to 20\% from one in some kinematic regions, but it is typically much smaller than that.\par

\subsection{Azimuthal angle resolution}
The finite resolution in the detector leads to systematic dilutions of the true asymmetries when extracting the asymmetry from the azimuthal dependence of the cross sections. From the simulation, \textit{e.g.}, the Collins angle ($\phi_{C} = \phi_{S} - \phi_{H}$) is calculated at both the detector-level and the associated particle-level. The difference of the reconstructed azimuthal angle of hadrons ($\delta \phi_{C}$ for the Collins effect) from the true value at the particle-level can be evaluated. The $\delta \phi_{C}$ distribution is convoluted with a unit sinusoidal distribution. The amplitude of the resulting curve is taken as the size of the dilution due to the finite azimuthal resolution, labeled as $f_{res}$. The final asymmetries and statistical uncertainties are corrected for the azimuthal dilutions by:
\begin{equation}
\label{equ:Azimuthal_Res_asym}
A_{pure} \rightarrow A_{pure,corr} = \frac{A_{pure}}{f_{res}}.
\end{equation}

The systematic uncertainty due to this correction is evaluated differently for the 2012 and 2015 data. For the 2012 data analysis, the correction is averaged over pion charge sign, so half of the difference between the correction for positively and negatively charged pions is used for the systematic uncertainty. In the 2015 data analysis, the difference between the primary 2015 simulation sample and the one that assumed the 2012 detector configuration is assigned as the systematic uncertainty.

\subsection{Non-uniform acceptance effects}\label{sec:leak_through}
The azimuthal modulations that appear in Eq.~(\ref{eq:crosssec}) are orthogonal. Therefore, in the limit of uniform detector acceptance, when the cross ratio in Eq.~(\ref{AN}) is binned in a particular combination of azimuthal angles, a sin($\phi$) fit will extract that specific modulation, and all other modulations will integrate out. The STAR detector has excellent azimuthal symmetry about the beam axis, allowing for the independent extraction of the Collins, Collins-like and inclusive jet asymmetries. However, any deviation from perfect azimuthal symmetry can introduce mixing of the modulations via coupling with a non-uniform acceptance. With this non-uniform acceptance, the components from other Fourier expansions can be modulated by the acceptance in a way that distorts the azimuthal dependence of the desired asymmetry.\par 

A data-driven approach is used to set an upper limit on the possible contamination to each of the three moments presented in this paper from the other two physics signals. This is done by artificially introducing a known asymmetry, $A_{in}$, associated with one physics modulation, then determining the impact on the other two physics modulations. The method has been used previously~\cite{ref:500GeVCollins}; the Collins asymmetry is used as an example in the following. To estimate the cross-talk from the inclusive jet asymmetry into the Collins asymmetry, two weights are constructed:
\begin{equation}
\begin{split}
& w_{0} = 1 + A_{in}\sin(\phi_{S}),\\
& w_{1} = 1 - A_{in}\sin(\phi_{S}),\\
\end{split}
\end{equation}
where $A_{in}$ is the known input asymmetry $A_{UT}^{\sin(\phi)}$. Each event is used twice, once with spin up and once with spin down. Events are given a weight $w_{0}$ when treated as spin up, and a weight $w_{1}$ when treated as spin down. The weighted events are fit to extract the induced Collins asymmetry. To estimate cross-talk from the Collins-like effect into the Collins asymmetry, the same procedure is applied, simply replacing the $\phi_{S}$ with $\phi_{CL} = \phi_{S} -2\phi_{H}$. The resulting amplitude ($p_{1}$) from the asymmetry fit determines the amount of cross-talk. The cross-talk systematic uncertainty is then estimated as:
\begin{equation}
\sigma_{cross} = \frac{\mathrm{Max}(|p_{1}|,\sigma_{p_{1}})}{A_{in}} \times \mathrm{Max}(|A_{meas}|, \sigma_{A_{meas}}),
\end{equation}
where $A_{meas}$ is the measured inclusive jet (Collins-like) asymmetry when estimating the inclusive jet (Collins-like) cross-talk into the Collins asymmetry. $p_{1}$ is the amplitude found in the cross-talk study, while $\sigma_{p_{1}}$ is the associated uncertainty.\par

For the contamination to the inclusive jet asymmetry by Collins or Collins-like effects, the weights are constructed by looping over all pions in the event and weighting by $z$ or $1 - z$ for the Collins or Collins-like case, respectively. The input asymmetries for the Collins effect have opposite sign for $\pi^{+}$ and $\pi^{-}$, while the same value is used for the Collins-like effect. Similar calculations are also made for each of the desired effects and their potential contamination.  Typically $\sigma_{cross}$ is less than 10\% of the statistical uncertainties.\par

\subsection{$A_{UT}^{\sin(\phi_S+\phi_H)}$ and $A_{UT}^{\sin(\phi_S+2\phi_H)}$}

As noted in Sec.\@ \ref{sec:AziMod}, the $A_{UT}^{\sin(\phi_S+\phi_H)}$ and $A_{UT}^{\sin(\phi_S+2\phi_H)}$ moments are expected to be negligible even in maximized scenarios \cite{ref:D_Alesio_2011}.  For this reason, all of the relevant corrections and systematic uncertainties in these two cases are not calculated.  However, the corresponding asymmetries are measured as functions of jet-$p_T$, and the results are found to be consistent with zero within the statistical uncertainties. For $x_F>0$, averaging over $p_T$, the asymmetries are:
$A_{UT}^{\sin(\phi_S+\phi_H)}(\pi^+)$ = -0.00011\,$\pm$\,0.00056 with $\chi^2$ = 9.55 for 9 degrees of freedom;
$A_{UT}^{\sin(\phi_S+\phi_H)}(\pi^-)$ = 0.00010\,$\pm$\,0.00058 with $\chi^2$ = 12.91 for 9 degrees of freedom;
$A_{UT}^{\sin(\phi_S+2\phi_H)}(\pi^+)$ = 0.00032\,$\pm$\,0.00062 with $\chi^2$ = 9.45 for 9 degrees of freedom; and
$A_{UT}^{\sin(\phi_S+2\phi_H)}(\pi^-)$ = -0.00037\,$\pm$\,0.00064 with $\chi^2$ = 3.39 for 9 degrees of freedom.

\begin{figure}
  \includegraphics[width=1\columnwidth]{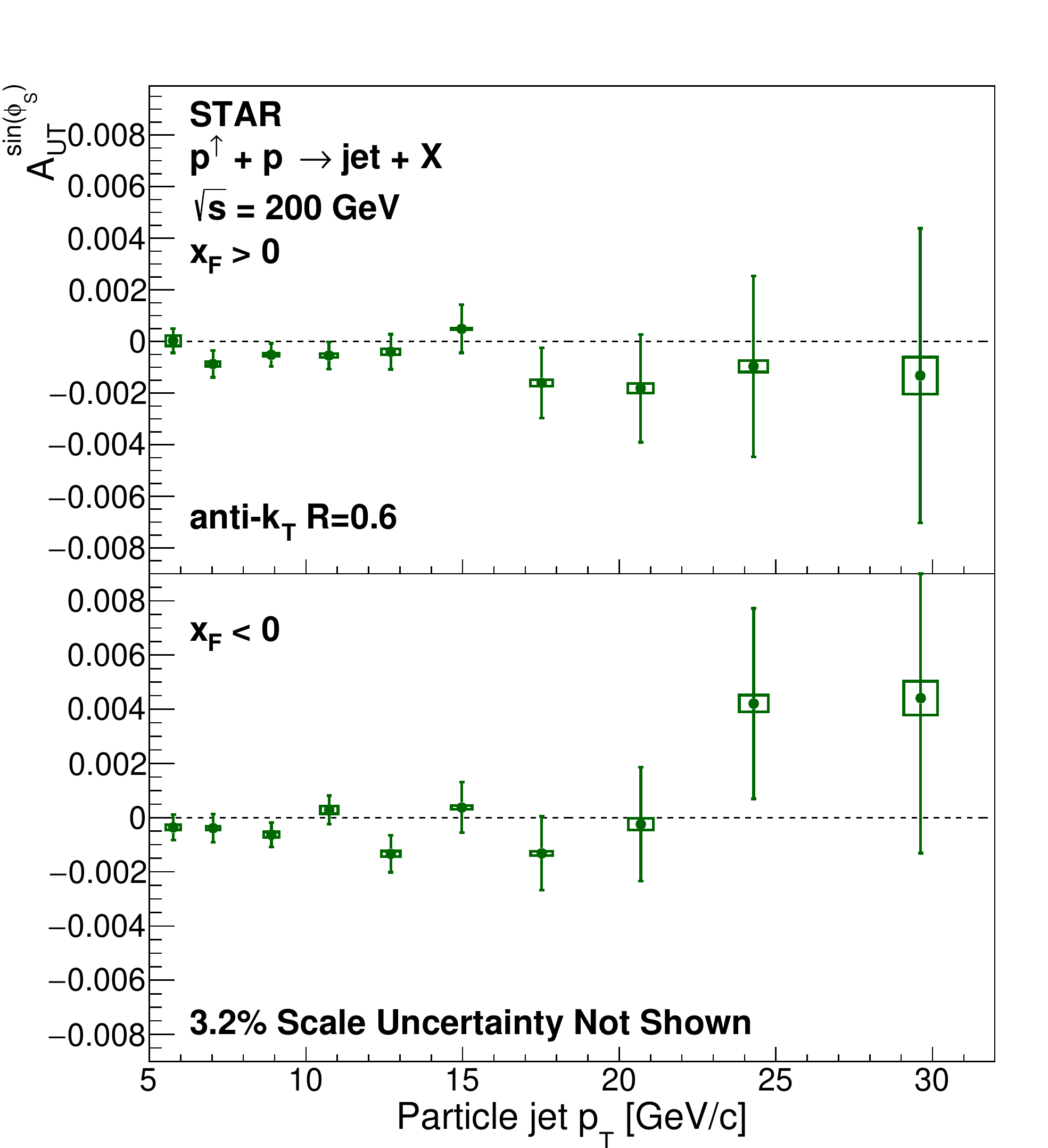}
  \caption{Inclusive jet asymmetries, $A_{UT}^{\sin(\phi_{S})}$, as a function of particle jet-$p_{T}$. The bars show the statistical uncertainties, while the size of the boxes represents the systematic uncertainties on $A_{UT}^{\sin(\phi_{S})}$ (vertical) and jet-$p_{T}$ (horizontal). The top panel shows results for jets that scatter forward relative to the polarized beam ($x_{F} > 0$), while the bottom panel shows jets that scatter backward to the polarized beam ($x_{F} < 0$). These results combine the 2012 and 2015 data.}
  \label{fig:sivers_final_inclusive}
\end{figure}

\begin{figure}
    \includegraphics[width=1\columnwidth]{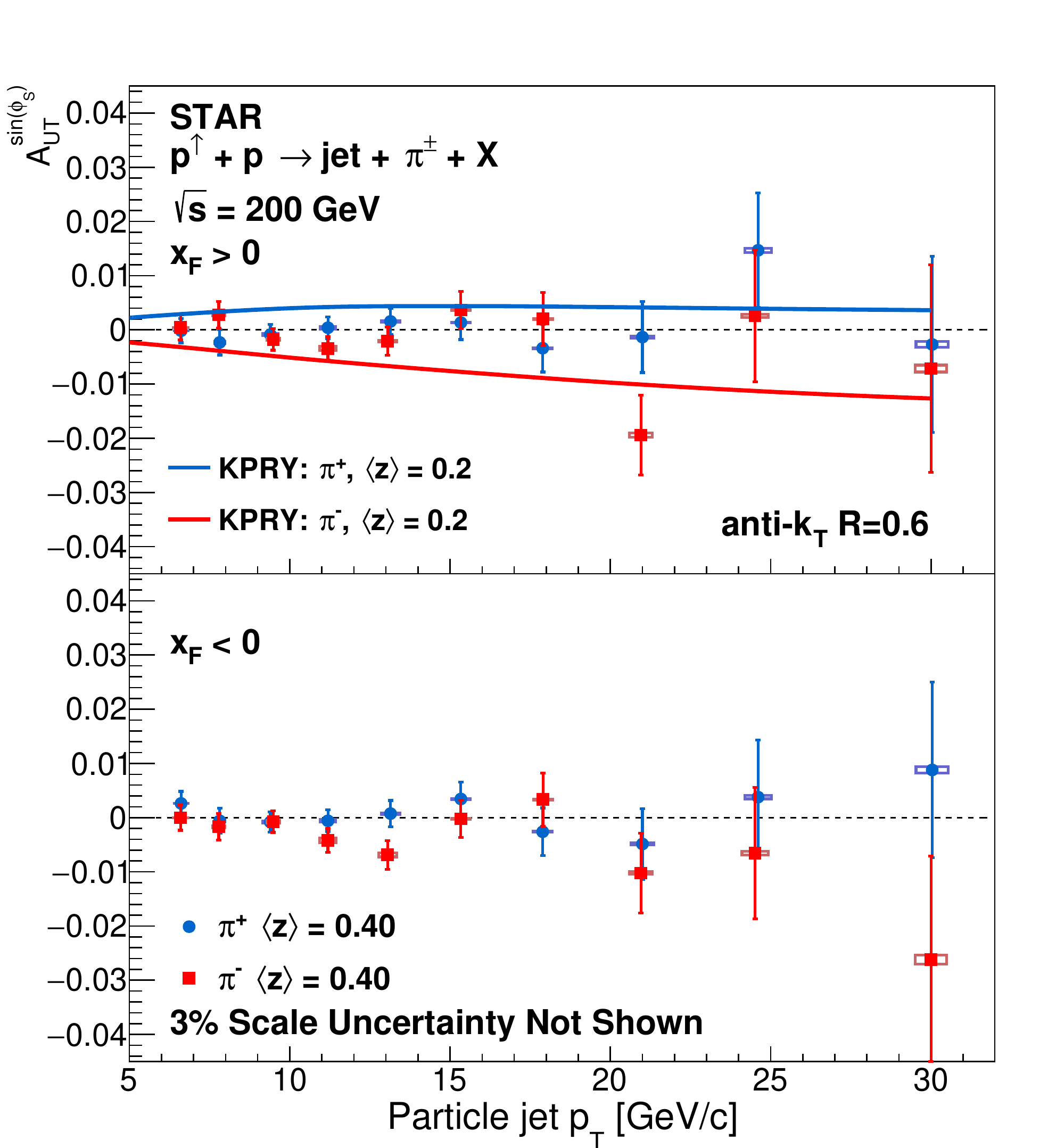}
    \caption{Hadron-tagged jet asymmetries, $A_{UT}^{\sin(\phi_{S})}$, as a function of particle jet-$p_{T}$ for jets that contain a charged pion with $z > 0.3$. The blue circles are for jets containing a high-$z$ $\pi^{+}$, while red squares are for jets containing a high-$z$ $\pi^{-}$. These results are from 2015 data. The asymmetries are shown in comparison with KPRY model calculations from Ref.~\cite{ref:Kang_2017b}. The theoretical calculations have the same colors as the data, and are calculated for a mean $z$ of 0.2.}
    \label{fig:figure_sivers_pi}
\end{figure}

\section{Final Results}\label{sec:Results}
The final asymmetries are shown as functions of the jet transverse momentum $p_T$, the hadron longitudinal momentum fraction $z$, and the hadron momentum transverse to the jet axis $j_{T}$, all of which have been corrected back to the particle-level. In each figure, the statistical uncertainties are shown with error bars, while the boxes on the data points show the systematic uncertainties. The heights of the uncertainty boxes represent the quadrature sum of the systematic uncertainties in $A_{UT}$ due to the contributions from underlying event dilutions, particle identifications, trigger bias, azimuthal resolutions and non-uniform acceptance. The widths of the uncertainty boxes represent the total systematic uncertainty associated with the jet or hadron energy scale as discussed in Sec.\@ \ref{subsec:kinematic_correction}. In the plots which show asymmetries for identified hadrons, the blue circles are for $\pi^{+}$, $K^{+}$ or $p$, while the red squares are for $\pi^{-}$, $K^{-}$ or $\bar{p}$.  Unless stated otherwise, the results from the 2012 and 2015 data analyses are combined in the following.\par

The results with jet-$p_{T}$ dependence are divided into two pseudorapidity ranges, one consisting of jets that scatter forward ($x_{F} > 0$) relative to the polarized beam, and the other for jets that scatter backward ($x_{F} < 0$) to the polarized beam. Positive $x_F$ jets are more likely to probe higher momentum fraction ($x$) partons and have both a larger quark jet fraction and a larger quark polarization transfer in the hard scattering. These considerations reverse for jets that scatter backward with respect to the polarized beam.  The latter are more likely to sample lower $x$ partons and have a larger gluon jet fraction and smaller quark polarization transfer. For the measurements involving multi-dimensional binning, due to the limited statistics, only the results for jets that scatter forward with respect to the polarized beam are presented here. The analogous results for jets that scatter backward with respect to the polarized beam are shown in the Appendix.

An overall vertical scale uncertainty of 3.2\% (3\% for the 2015 data and 3.5\% for the 2012 data) from the beam polarization uncertainty is not shown.\par

\subsection{Inclusive jet asymmetries}\label{subsec:inclusive_asym}

Figure~\ref{fig:sivers_final_inclusive} shows the inclusive jet asymmetry ($A_{UT}^{\mathrm{sin}(\phi_{S})}$) with jet-$p_{T}$ dependence. At midrapidity, this value is expected to be dominated by the gluon Sivers function via the twist-3 correlators.  The measured asymmetries are consistent with zero within uncertainties, similar to the previous STAR measurements in $pp$ collisions at 200 GeV~\cite{ref:StarSivers2006} and 500 GeV~\cite{ref:500GeVCollins}.  However, the uncertainties for the present results are an order of magnitude smaller than those for the previous 200 GeV measurement \cite{ref:StarSivers2006}.  They are also a factor of four smaller than those for the previous 500 GeV measurement \cite{ref:500GeVCollins} when compared within the common $x_T = 2 p_T/\sqrt{s}$ range, $0.06 < x_T < 0.2$.  Thus, they should further constrain the midrapidity twist-3 models.

Figure~\ref{fig:figure_sivers_pi} shows the first measurement of $A_{UT}^{\mathrm{sin}(\phi_{S})}$ for jets that contain a charged pion with a high longitudinal momentum fraction $z > 0.3$. In this way, the $u$ (for $\pi^{+}$) and $d$ (for $\pi^{-}$) quark fractions are enhanced in the measurement \cite{ref:D_Alesio_2011,ref:Gamberg_2013}. Theoretical expectations from the KPRY model~\cite{ref:Kang_2017b} are also shown in Fig.~\ref{fig:figure_sivers_pi}. However, the mean $z$ range is 0.2 for this theoretical calculation, which is well below the range of the data.\par

\subsection{Collins-like asymmetries}

\begin{figure}[!ht]
    \includegraphics[width=1.0\columnwidth]{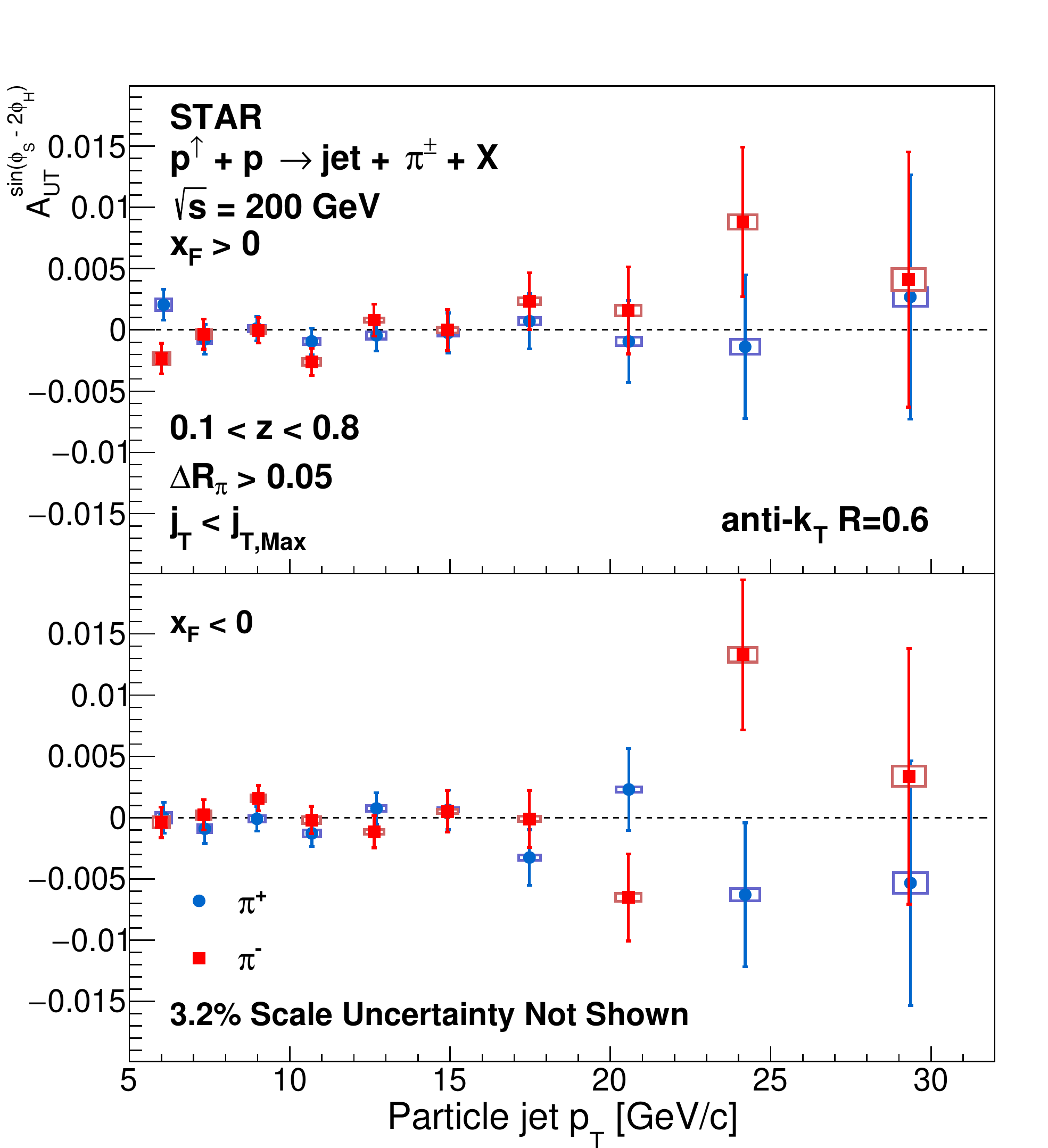}
    \caption{Collins-like asymmetries, $A_{UT}^{\sin(\phi_{S}-2\phi_{H})}$, as a function of particle jet-$p_{T}$. The bars show the statistical uncertainties, while the size of the boxes represents the systematic uncertainties on $A_{UT}^{\sin(\phi_{S}-2\phi_{H})}$ (vertical) and jet-$p_{T}$ (horizontal). The top panel shows results for jets that scatter forward relative to the polarized beam ($x_{F} > 0$), while the bottom panel shows jets that scatter backward with respect to the polarized beam ($x_{F} < 0$).}
    \label{fig:figure_CollinsLike_vs_jetPt}
\end{figure}

\begin{figure*}
    \includegraphics[width=2.0\columnwidth]{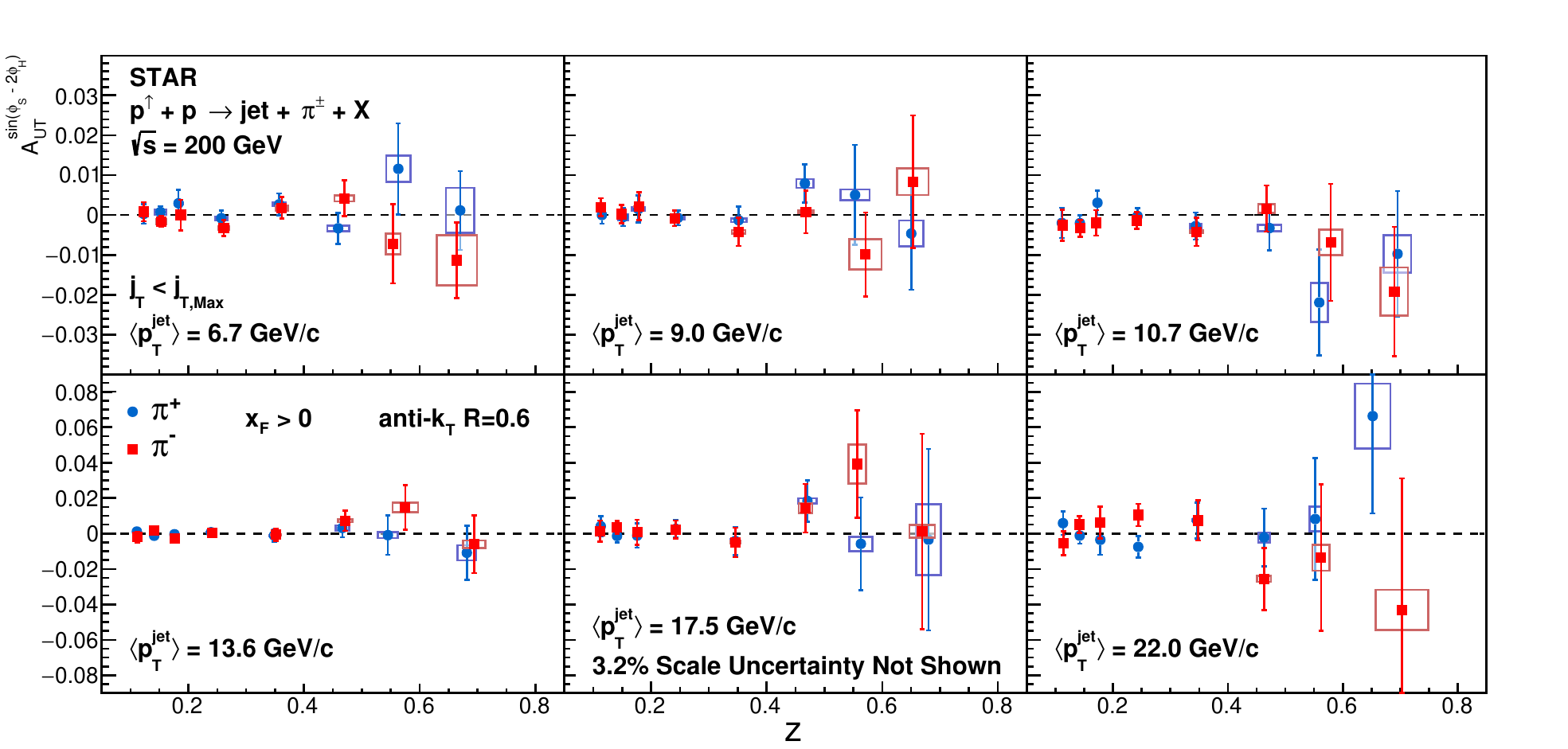}
    \caption{Collins-like asymmetries, $A_{UT}^{\sin(\phi_{S}-2\phi_{H})}$, as a function of the charged pion's longitudinal momentum fraction, $z$, in different jet-$p_{T}$ bins. The bars show the statistical uncertainties, while the size of the boxes represents the systematic uncertainties on $A_{UT}^{\sin(\phi_{S}-2\phi_{H})}$ (vertical) and hadron-$z$ (horizontal).}
    \label{fig:figure_CollinsLike_vs_z}
\end{figure*}

Figure~\ref{fig:figure_CollinsLike_vs_jetPt} shows the Collins-like asymmetries with the modulation of $A_{UT}^{\sin(\phi_{S}-2\phi_{H})}$ for charged pions within jets as a function of jet-$p_T$, and Fig.~\ref{fig:figure_CollinsLike_vs_z} shows the hadron-$z$ dependence in six different jet-$p_{T}$ bins. Both the jet-$p_{T}$ and hadron-$z$ values are corrected back to the particle-level. These results represent the first measurement of the Collins-like moment in $\sqrt{s}$ = 200 GeV polarized proton-proton collisions. Across the covered kinematic range, there is no observed significant asymmetry for either charge state.\par

The Collins-like asymmetry is sensitive to the distribution of linearly polarized gluons in transversely polarized protons. Therefore the asymmetry is expected to be largest at low values of jet-$p_{T}$ where gluons constitute the majority of the subprocess fraction. However, there is no significant asymmetry observed here. This result is consistent with the previous STAR measurement of the Collins-like effect in $pp$ collisions at $\sqrt{s}$ = 500 GeV from data collected in 2011 \cite{ref:500GeVCollins}, but with significantly increased statistical precision within the common $x_T$ range. These data, in conjunction with the results in Ref.~\cite{ref:500GeVCollins}, will help to constrain model calculations, such as those in Ref.~\cite{ref:Gamberg_2013}.\par

\subsection{Collins asymmetries}\label{sec:Collins_results}
\subsubsection{Pion Collins asymmetries} 
\begin{figure}[!ht]
    \includegraphics[width=1.0\columnwidth]{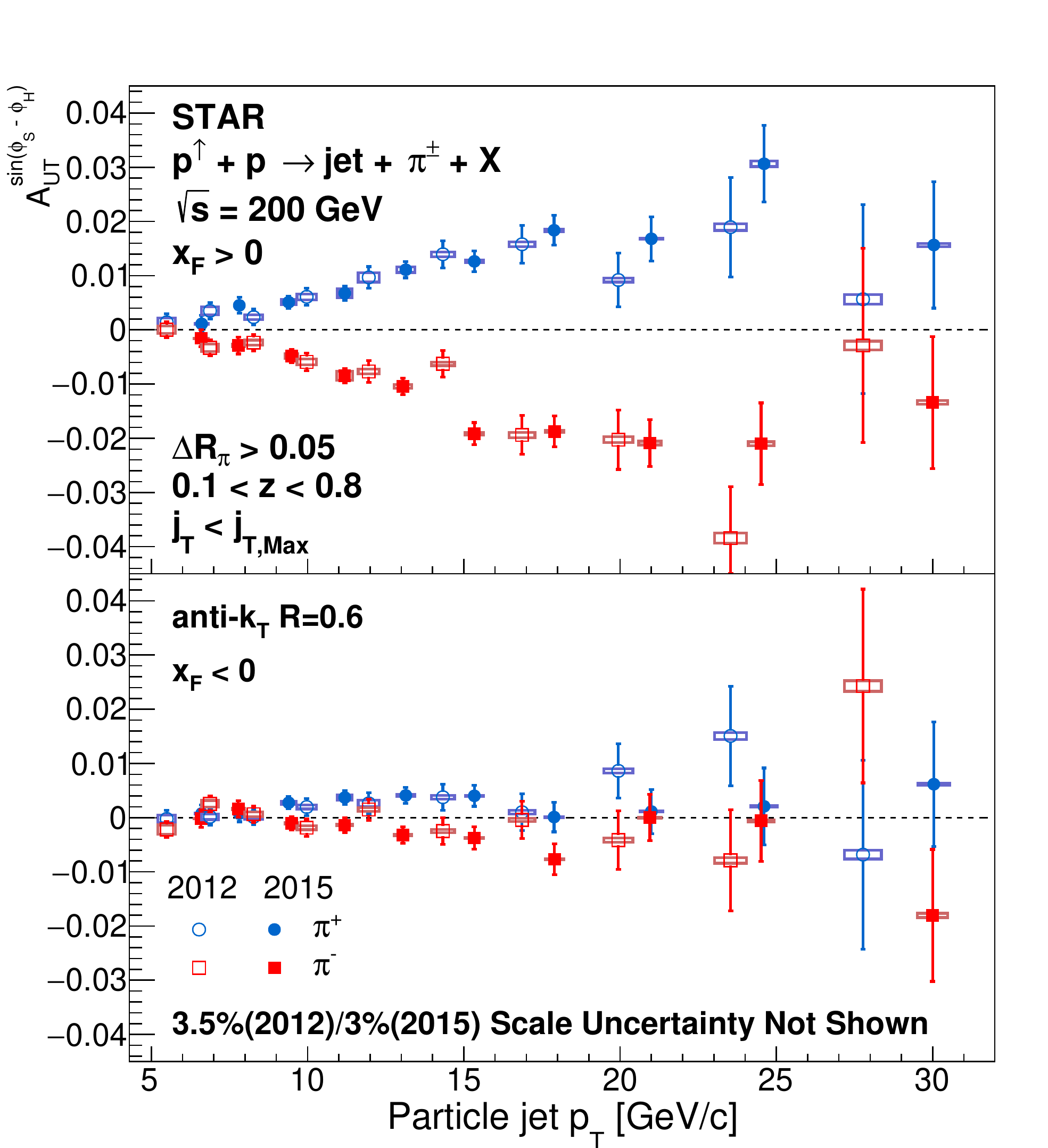}
    \caption{Collins asymmetries, $A_{UT}^{\sin(\phi_{S}-\phi_{H})}$, as a function of particle jet-$p_{T}$ separately for the 2012 and 2015 data. The bars show the statistical uncertainties, while the size of the boxes represents the systematic uncertainties on $A_{UT}^{\sin(\phi_{S}-\phi_{H})}$ (vertical) and jet-$p_{T}$ (horizontal). The top panel shows the results for jets that scatter forward relative to the polarized beam ($x_{F} > 0$), while the bottom panel shows jets that scatter backward to the polarized beam ($x_{F} < 0$).}
    \label{fig:figure_Collins_vs_jetPt_2012_2015}
\end{figure}

Figure~\ref{fig:figure_Collins_vs_jetPt_2012_2015} shows the results of the Collins asymmetry, $A_{UT}^{\sin(\phi_{S}-\phi_{H})}$, separately for the two data sets, 2012 and 2015.  Note that the horizontal offset between the results from the two years is not artificial.  It arises from the different treatment of underlying event effects at the detector-level in the two analyses which was described in Sec. \ref{subsec:UE_corrs}.  The results from the independent analyses of data from two different RHIC running periods are in good agreement.  The 2012 and 2015 results are combined in the subsequent plots.

\begin{figure}[!ht]
    \includegraphics[width=1.0\columnwidth]{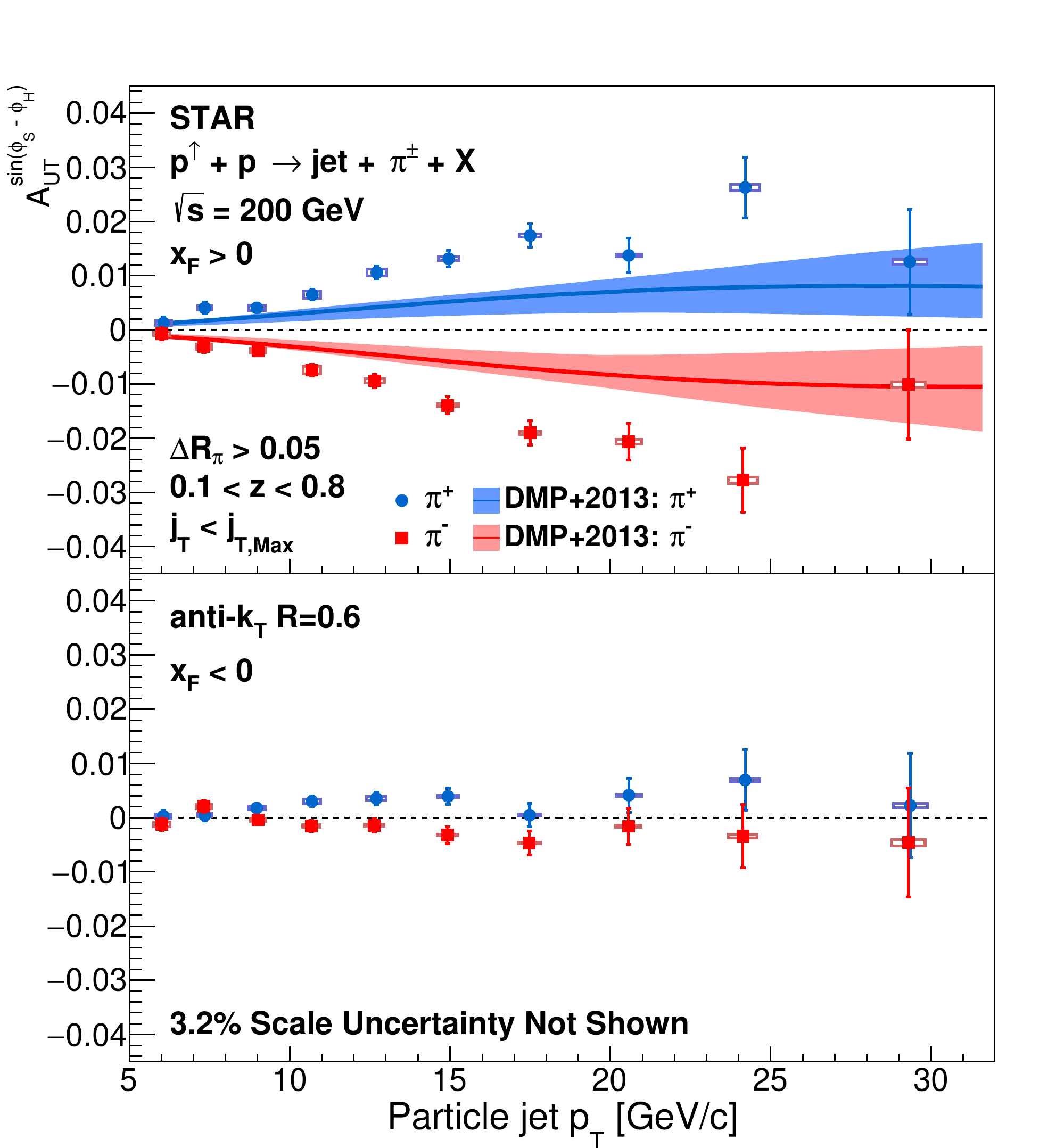}
    \caption{Collins asymmetries, $A_{UT}^{\sin(\phi_{S}-\phi_{H})}$, as a function of particle jet-$p_{T}$. The bars show the statistical uncertainties, while the size of the boxes represents the systematic uncertainties on $A_{UT}^{\sin(\phi_{S}-\phi_{H})}$ (vertical) and jet-$p_{T}$ (horizontal). The top panel shows the results for jets that scatter forward relative to the polarized beam ($x_{F} > 0$), while the bottom panel shows jets that scatter backward to the polarized beam ($x_{F} < 0$). The asymmetries are shown in comparison with DMP+2013 model calculations from Ref.~\cite{ref:D_Alesio_2017}.}
    \label{fig:figure_Collins_vs_jetPt}
\end{figure}

\begin{figure*}
\centering
    \includegraphics[width=2.0\columnwidth]{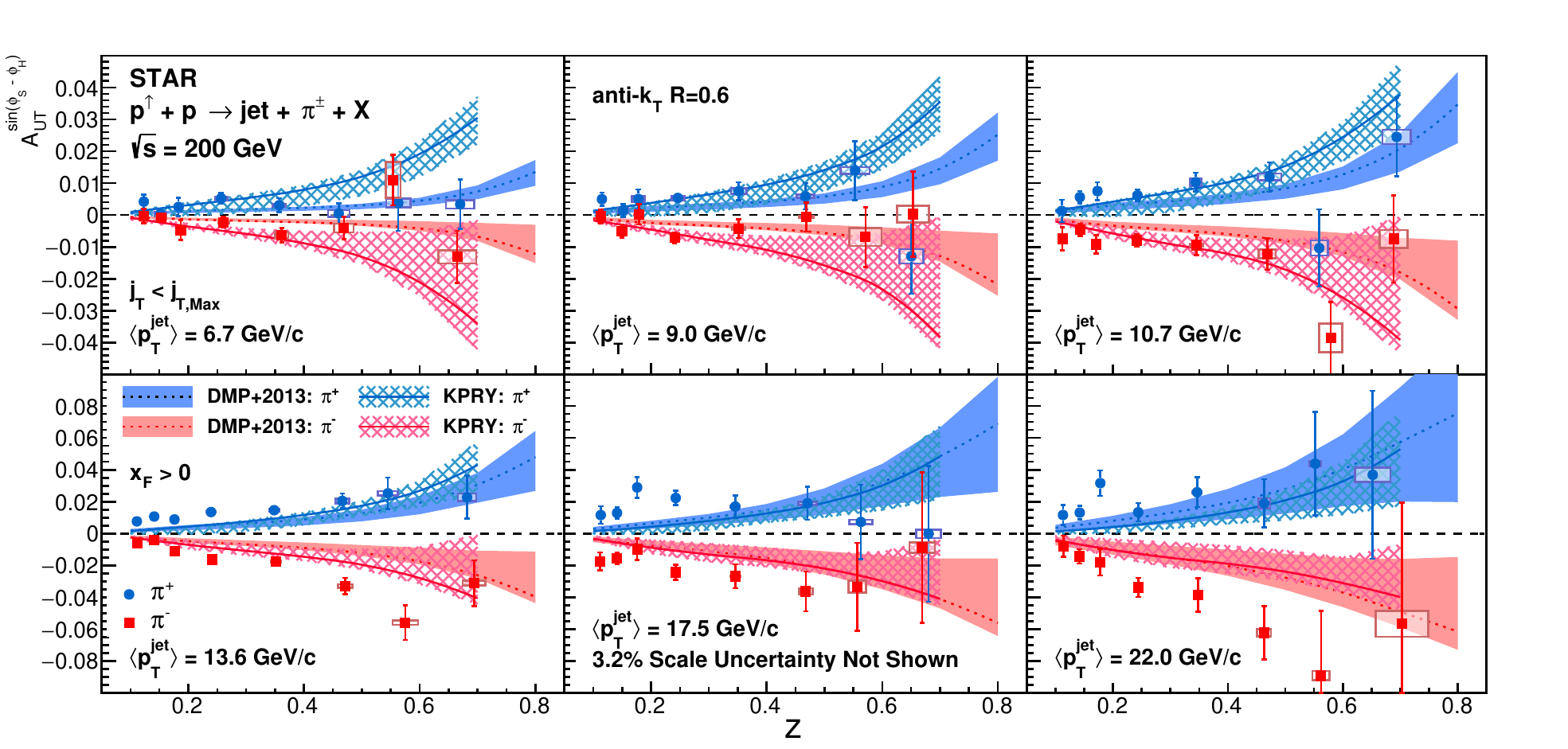}
    \caption{Collins asymmetries, $A_{UT}^{\sin(\phi_{S}-\phi_{H})}$, as a function of the charged pion's longitudinal momentum fraction, $z$, in different jet-$p_{T}$ bins. The bars show the statistical uncertainties, while the size of the boxes represents the systematic uncertainties on $A_{UT}^{\sin(\phi_{S}-\phi_{H})}$ (vertical) and hadron-$z$ (horizontal). The asymmetries are shown in comparison to calculations with the DMP+2013 model from Ref.~\cite{ref:D_Alesio_2017} and the KPRY model from Ref.~\cite{ref:Kang_2017b}.}
    \label{fig:figure_Collins_vs_z}
\end{figure*}

\begin{figure*}
\centering
    \includegraphics[width=2.0\columnwidth]{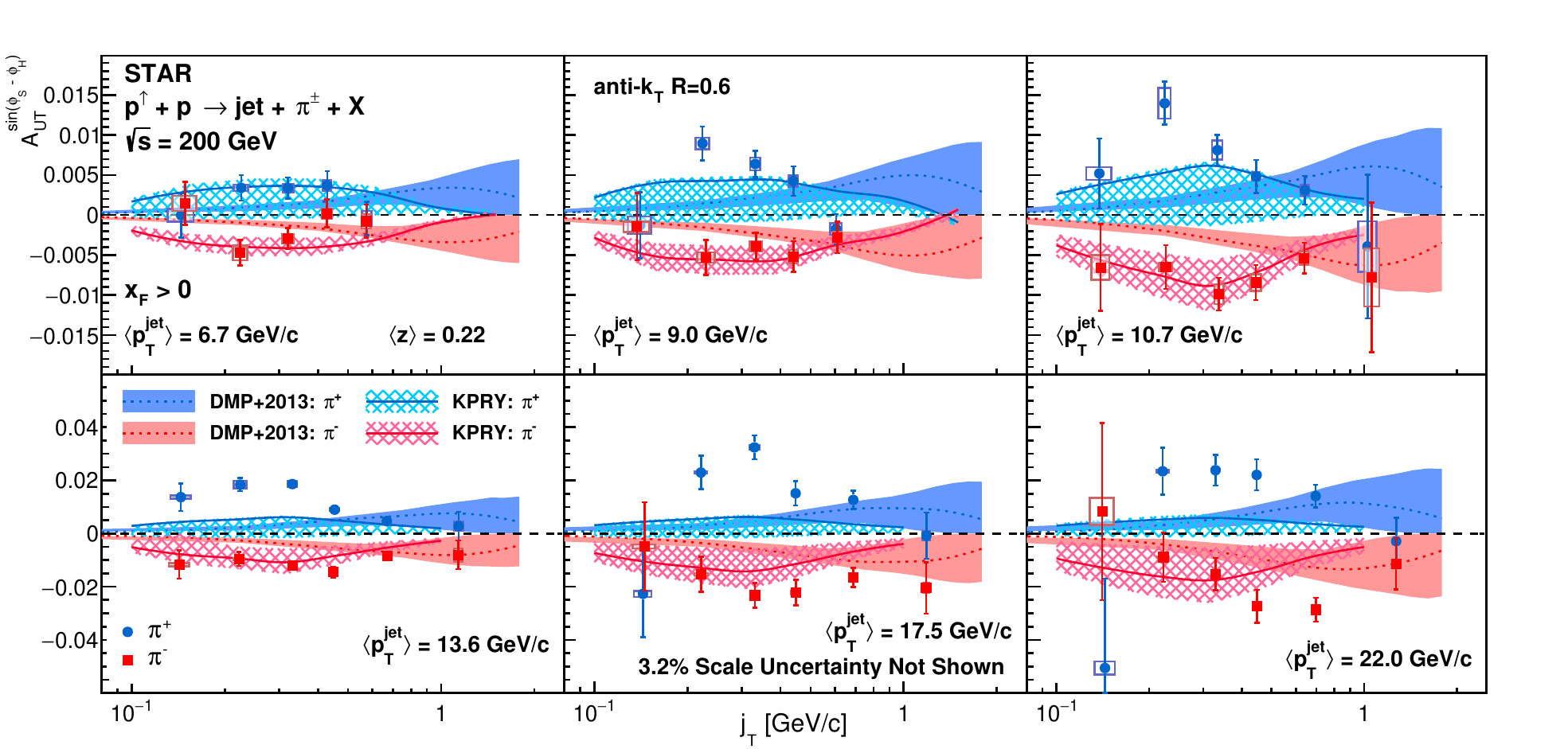}
    \caption{Collins asymmetries, $A_{UT}^{\sin(\phi_{S}-\phi_{H})}$, as a function of the charged pion's momentum transverse to the jet axis, $j_{T}$, in different jet-$p_{T}$ bins. The bars show the statistical uncertainties, while the size of the boxes represents the systematic uncertainties on $A_{UT}^{\sin(\phi_{S}-\phi_{H})}$ (vertical) and hadron-$j_{T}$ (horizontal). The asymmetries are shown in comparison to calculations with the DMP+2013 model from Ref.~\cite{ref:D_Alesio_2017} and the KPRY model from Ref.~\cite{ref:Kang_2017b}.}
    \label{fig:figure_Collins_vs_jT_jetPt}
\end{figure*}

The Collins asymmetries $A_{UT}^{\sin(\phi_{S}-\phi_{H})}$ for charged pions within jets are presented in Figs.~\ref{fig:figure_Collins_vs_jetPt}$-$\ref{fig:figure_Collins_vs_jT_jetPt}. A scheme of various two-dimensional binning is employed to elucidate fine details that can be washed out when variables are integrated over, as in Fig.~\ref{fig:figure_Collins_vs_jetPt}. In contrast to the Collins-like
asymmetry, the Collins asymmetry exists for quarks only. Therefore the signal is expected to be largest for high jet-$p_{T}$ where quark events are far more likely.\par

\begin{figure}[!ht]
    \includegraphics[width=1.0\columnwidth]{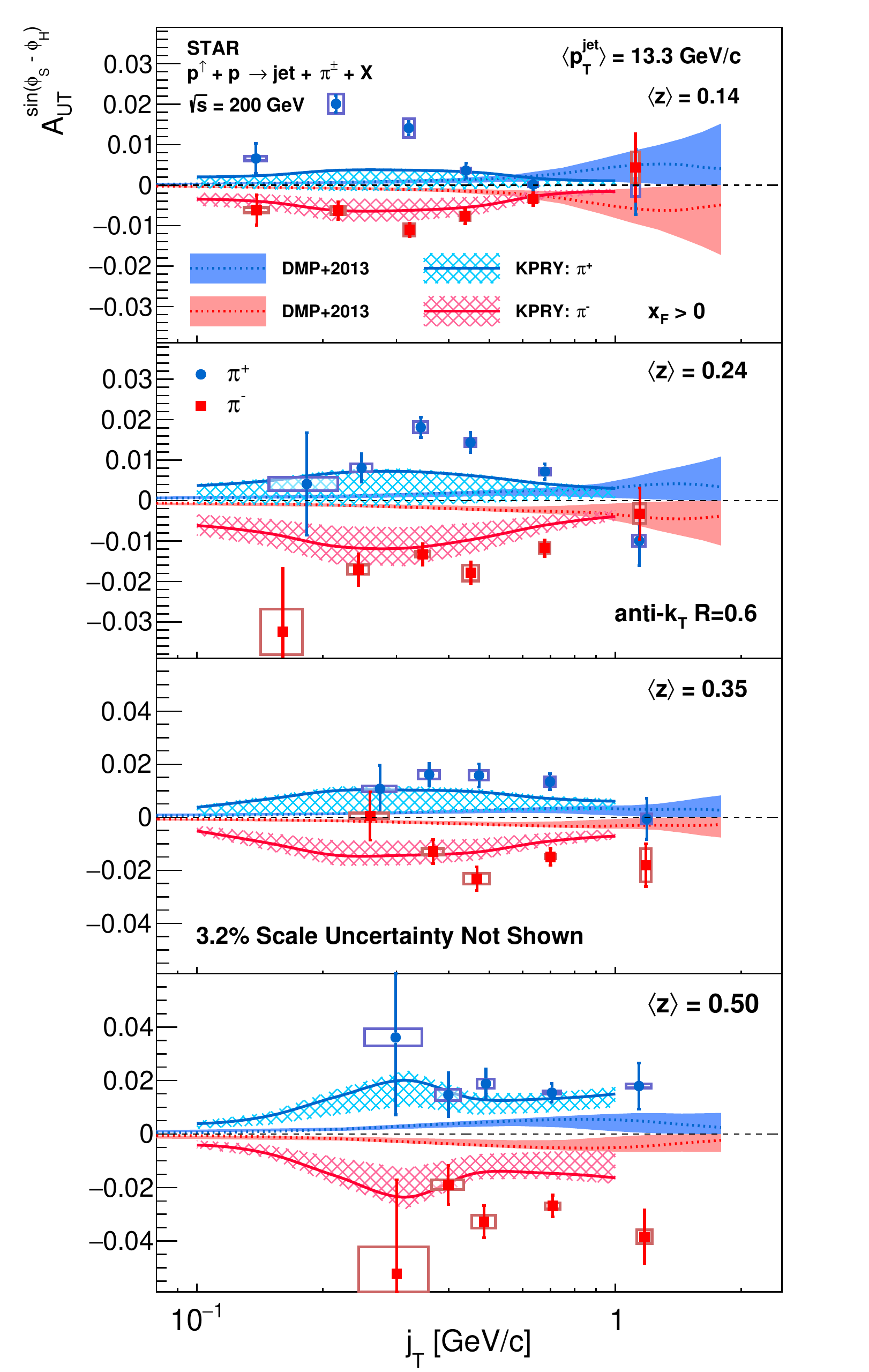}
    \caption{Collins asymmetries, $A_{UT}^{\sin(\phi_{S}-\phi_{H})}$, as a function of the charged pion's momentum transverse to the jet axis, $j_{T}$, in different hadron longitudinal momentum fraction $z$ bins, integrated over detector jet-$p_T > 9.9$ GeV/$c$. The bars show the statistical uncertainties, while the size of the boxes represents the systematic uncertainties on $A_{UT}^{\sin(\phi_{S}-\phi_{H})}$ (vertical) and hadron-$j_{T}$ (horizontal). The asymmetries are shown in comparison to calculations with the DMP+2013 model from Ref.~\cite{ref:D_Alesio_2017} and the KPRY model from Ref.~\cite{ref:Kang_2017b}.}
    \label{fig:figure_Collins_vs_jT_zbin}
\end{figure}

Figure~\ref{fig:figure_Collins_vs_jetPt} shows the Collins asymmetries for pions as a function of jet-$p_{T}$, separately for $x_{F} > 0$ and $x_F < 0$.  The asymmetries are small at lower values of jet-$p_T$, and then increase to large values for $x_F > 0$ as jet-$p_{T}$ gets larger. The $\pi^{+}$ asymmetries are positive, while the $\pi^{-}$ asymmetries are negative, with similar magnitudes.  These trends also extend into the $x_F < 0$ region, though with much smaller magnitudes than found for $x_F > 0$. These results are integrated over a wide range of $z$ and $j_{T}$ in order to provide sensitivity to the collinear transversity distributions. The figure also shows theoretical expectations based on the DMP+2013 model from Ref.~\cite{ref:D_Alesio_2017}. The DMP+2013 model uses the leading order TMD approach, and is based on a fit to transversity and Collins fragmentation function measurements from SIDIS and $e^{+}e^{-}$ processes~\cite{ref:Anselmino_2015}. The presented error bands for the theoretical calculations represent the statistical uncertainties and correspond to a 95.45\% Confidence Level (CL) during their fits. The data presented here are larger in magnitude than the expectations from this model.\par

The mean $\langle z \rangle$ for these Collins asymmetry measurements varies from 0.23 at low jet-$p_T$ to 0.19 at high jet-$p_T$. The mean $j_{T}$ increases from 0.3 to 0.6 GeV/$c$ as jet-$p_{T}$ increases. For $x_F>0$, the mean $\langle x \rangle$ from the polarized proton also increases from low to high jet-$p_{T}$, and ranges from 0.1 to 0.4.\par

The functional dependence of the asymmetry is explored in greater detail in the following plots, {\it e.g.}, Fig.~\ref{fig:figure_Collins_vs_z},~\ref{fig:figure_Collins_vs_jT_jetPt},~\ref{fig:figure_Collins_vs_jT_zbin}). Figure~\ref{fig:figure_Collins_vs_z} shows the $z$ dependence of the Collins asymmetries separately for six different jet-$p_{T}$ bins.  The asymmetries increase with $z$ and jet-$p_{T}$. Both the DMP+2013 model~\cite{ref:D_Alesio_2017} and KPRY model~\cite{ref:Kang_2017b} expectations are presented. The KPRY model is also calculated based on a global analysis of SIDIS and $e^{+}e^{-}$ data~\cite{ref:Kang_2016}, and treats TMD evolution up to the next-to-leading logarithmic effects. The presented error bands are based on the uncertainties from the quark transversity distributions and the Collins fragmentation functions used in the model calculation. At low jet-$p_T$, the KPRY model predicts asymmetries which are larger than the DMP+2013 model and generally consistent with the measurements.  At high jet-$p_T$, the two models predict similar asymmetries, both of which undershoot the majority of the data. Both groups emphasize that the transverse momentum dependences of the unpolarized and the Collins fragmentation functions are not well understood.  This may be the main reason for the discrepancies.\par

Figures~\ref{fig:figure_Collins_vs_jT_jetPt} and ~\ref{fig:figure_Collins_vs_jT_zbin} show the $j_T$ dependence of the Collins asymmetry in six jet-$p_{T}$ bins and four hadron-$z$ bins, respectively.  For Fig.\@ \ref{fig:figure_Collins_vs_jT_zbin}, detector jet-$p_T$ is integrated over $p_T >9.9$ GeV/$c$ where significant Collins asymmetries are seen in Figs.\@ \ref{fig:figure_Collins_vs_jetPt} and \ref{fig:figure_Collins_vs_z}. DMP+2013 model~\cite{ref:D_Alesio_2017} and KPRY model~\cite{ref:Kang_2017b} expectations are also presented. Overall, the results favor the KPRY model.  However, significant discrepancies exist between the data and both model calculations.  Notably, the observed $\pi^+$ asymmetries at low-$z$ or high-jet-$p_T$ are consistently larger than predicted by either model.\par

\begin{figure}[!hbt]
    \includegraphics[width=1.0\columnwidth]{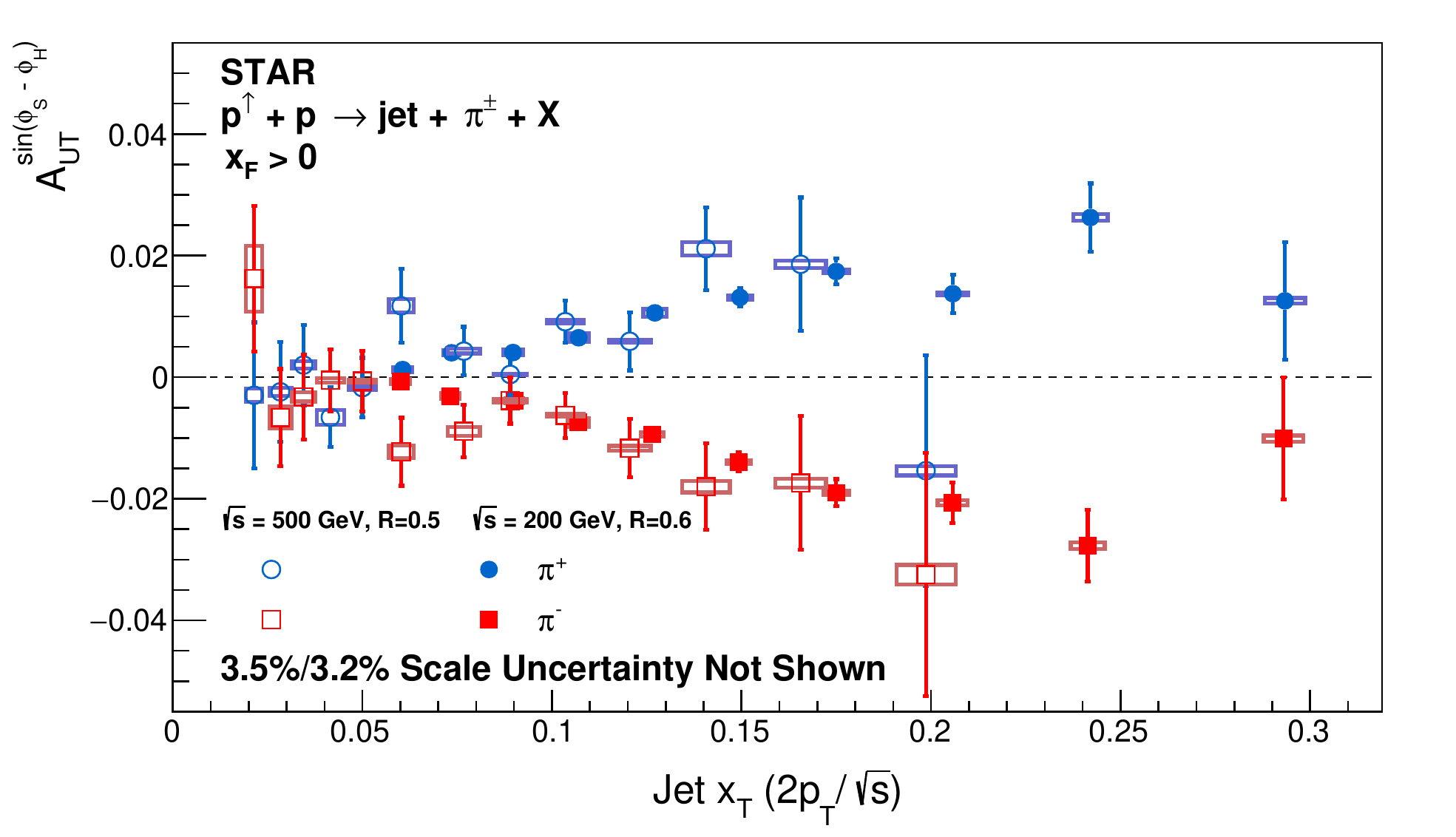}
    \caption{Collins asymmetries, $A_{UT}^{\sin(\phi_{S}-\phi_{H})}$, as a function of particle jet $x_{T}~(= 2 p_T/\sqrt{s}$). The solid points show the results from this analysis of $\sqrt{s} = 200$ GeV $pp$ collisions, while the open points show previous STAR results for $\sqrt{s} = 500$ GeV $pp$ collisions from data recorded during 2011~\cite{ref:500GeVCollins}. The 200 GeV results utilize a jet radius of 0.6, whereas the 500 GeV results utilized a jet radius of 0.5.}
    \label{fig:figure_Collins_vs_2011posxT}
\end{figure}

\begin{figure}[!hbt]
    \includegraphics[width=1.0\columnwidth]{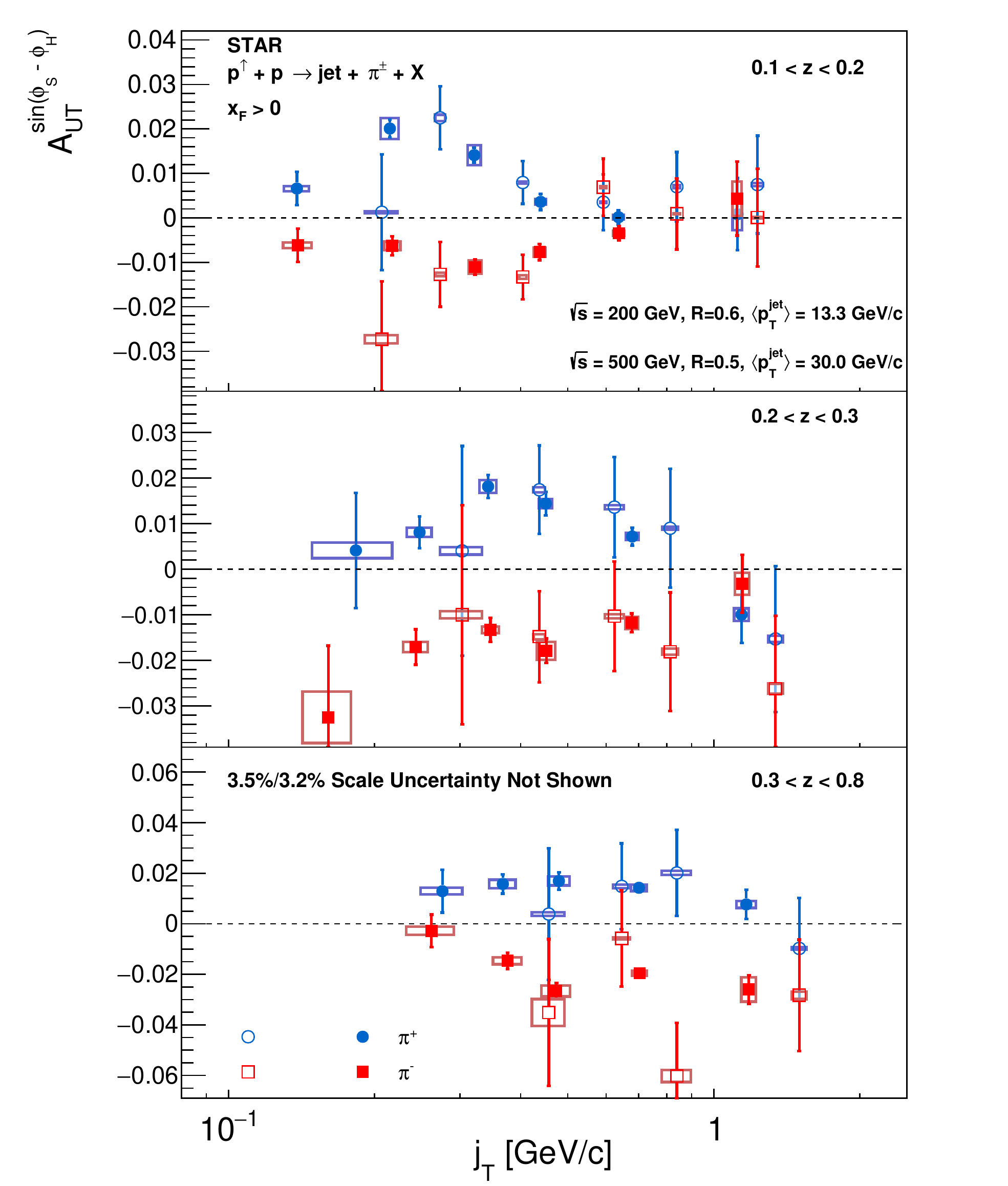}
    \caption{Collins asymmetries, $A_{UT}^{\sin(\phi_{S}-\phi_{H})}$, as a function of the charged pion momentum transverse to the jet axis, $j_{T}$, in different hadron longitudinal momentum fraction $z$ bins. The solid points show the results from this analysis of $\sqrt{s} = 200$ GeV $pp$ collisions, while the open points show previous STAR results for $\sqrt{s} = 500$ GeV $pp$ collisions from data recorded during 2011~\cite{ref:500GeVCollins}.}
    \label{fig:figure_Collins_vs_jT_zbin_with2011}
\end{figure}

\subsubsection{Comparison of asymmetries in 200 and 500 GeV collisions}
Figure~\ref{fig:figure_Collins_vs_2011posxT} shows the comparison of the new Collins asymmetry results at $\sqrt{s}$ = 200 GeV with the published $\sqrt{s} =$ 500 GeV measurements from the 2011 RHIC running period~\cite{ref:500GeVCollins} as functions of jet $x_{T} = 2p_{T}/\sqrt{s}$, which has corrected back to particle level. Jets are reconstructed using the same algorithm at 500 GeV with a radius of 0.5. The jet $x_{T}$ dependence of the 500 GeV results was separated into three different $z$ bins in the previous paper. The asymmetries are calculated as weighted means in order to provide similar kinematic coverage as in this analysis. The measured asymmetries agree with each other in the overlap region $0.06 < x_{T} < 0.2$, even though the corresponding $Q^2$ values differ by a factor of about six.  This indicates that the Collins asymmetry has at most a weak energy dependence in hadronic collisions.\par

Figure~\ref{fig:figure_Collins_vs_jT_zbin_with2011} shows the comparison of these new Collins asymmetry results at $\sqrt{s}$ = 200 GeV with the published $\sqrt{s}$ = 500 GeV measurements from the 2011 RHIC running period~\cite{ref:500GeVCollins} as a functions of pion $j_{T}$ for three different $z$ ranges. The asymmetries are plotted at similar values of $x_{T}$.  Once again, the asymmetries at the two collision energies agree, indicating that the Collins effect has at most a weak energy dependence in hadronic collisions.  By integrating over wide ranges of jet $x_T$, Figs.~\ref{fig:figure_Collins_vs_jT_zbin} and \ref{fig:figure_Collins_vs_jT_zbin_with2011} provide a detailed view of the kinematic dependence of the Collins fragmentation function, without the complication of being convoluted with the TMD transversity distribution~\cite{ref:Kang_2017a}. Notably, Figs.~\ref{fig:figure_Collins_vs_jT_zbin} and \ref{fig:figure_Collins_vs_jT_zbin_with2011} indicate that the $z$ and $j_T$ dependences of the Collins FF are not separable, in contrast to a common assumption that has been built into most previous global analyses of Collins asymmetry data. The current results will give the first input to the theory community to better understand how this relationship affects factorization models and global analyses.\par

\subsubsection{Kaon and proton Collins asymmetries}\label{subsubsec:K_p_Collins}
\begin{figure*}
  \centering
    \includegraphics[width=2\columnwidth]{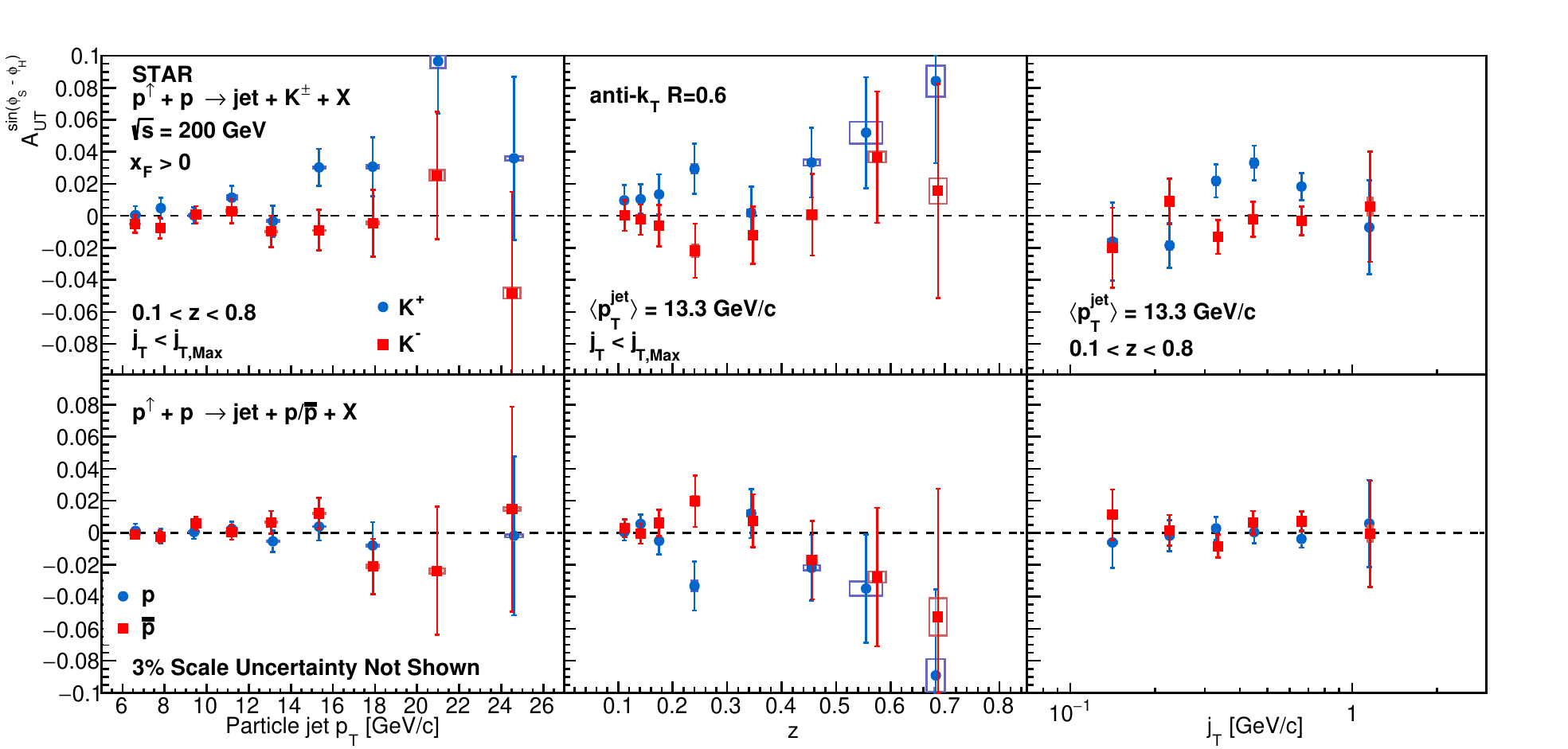}
    \caption{Collins asymmetries, $A_{UT}^{\sin(\phi_{S}-\phi_{H})}$, as a function of particle jet-$p_{T}$, hadron-$z$, and hadron-$j_{T}$ for charged kaons (upper panels) and protons (lower panels) inside jets.  In both cases, the $p_T$ dependence is shown integrated over the full ranges of $z$ and $j_T$, while the $z$ and $j_T$ dependences are shown integrated over detector jet-$p_T > 9.9$ GeV/$c$. The bars show the statistical uncertainties, while the size of the boxes represent the systematic uncertainties.}
    \label{fig:figure_Collins_Kaon_vs_jetPt}
\end{figure*}

The first measurement of the Collins asymmetries for charged kaons inside jets in $pp$ collisions are presented in the upper panels of Fig.~\ref{fig:figure_Collins_Kaon_vs_jetPt}. These results are  plotted with jet-$p_{T}$, hadron-$z$, and hadron-$j_{T}$ dependence from left to right panels.  The jet-$p_T$ dependence is shown integrated over the full ranges of $z$ and $j_T$, while the $z$ and $j_T$ dependences are shown integrated over detector jet-$p_T > 9.9$ GeV/$c$. Due to the limited statistics, the results are not further divided into multi-dimensional bins. The asymmetries for $K^{+}$, which like $\pi^{+}$ have a contribution from favored fragmentation of $u$ quarks, are similar in magnitude to those for $\pi^{+}$ in Fig.~\ref{fig:figure_Collins_vs_jetPt}~\ref{fig:figure_Collins_vs_z}, ~\ref{fig:figure_Collins_vs_jT_jetPt}. In contrast, the asymmetries for $K^{-}$, which can only come from unfavored fragmentation, are consistent with zero at the one sigma level. These trends are similar to those found in SIDIS by HERMES~\cite{ref:Airapetian_2010, ref:Airapetian_2020} and COMPASS~\cite{ref:Adolph_2015}, and provide additional insight into the dynamical origins of the Collins fragmentation function.\par

The first measurements of the Collins azimuthal asymmetries for protons inside jets are also presented here, as shown in the lower panels of Fig.~\ref{fig:figure_Collins_Kaon_vs_jetPt}. Similar to the kaon results, they are also plotted with jet-$p_{T}$, hadron-$z$, and hadron-$j_{T}$ dependence in three panels. Fragmentation into protons is much more complicated than into mesons~\cite{ref:PhysRevD.76.074033}, and is not expected to produce Collins asymmetries. Compared to a recent result released by HERMES~\cite{ref:Airapetian_2020}, where they concluded the asymmetries were mostly negative for protons and zero for anti-protons, the results presented here are all consistent with zero at the one sigma level.\par

\section{Summary}\label{sec:Conclusion}
In summary, new measurements of the transverse single-spin asymmetries for inclusive jet and identified hadron in jet production at midrapidity from transversely polarized $pp$ collisions at $\sqrt{s}$ = 200 GeV, based on data recorded by STAR during the 2012 and 2015 RHIC running periods, are presented. The inclusive jet asymmetry measurements are found to be consistent with zero, including $A_N$ for inclusive jets and $A_N$ for jets containing charged pions carrying longitudinal momentum fraction $z > 0.3$.  These results will provide more stringent constraints in future fits of the twist-3 analogs of the Sivers effect.

Collins-like asymmetries for charged pions within jets are found to be very small and consistent with zero at current precision. This allows new, more precise upper limits on linearly polarized gluons in transversely polarized protons.

Identified hadron in jet asymmetry measurements of the Collins asymmetry for charged pions, kaons and protons are presented for jets that scatter both forward and backward relative to the polarized beam. The asymmetries are large, opposite in sign, and have similar magnitude for $\pi^{+}$ and $\pi^{-}$ in jets with $x_{F} > 0$. The pion asymmetries in $\sqrt{s}$ = 200 GeV $pp$ collisions agree with previous measurements of the Collins asymmetries in $\sqrt{s} = 500$ GeV $pp$ collisions, indicating that the Collins effect has at most a weak energy dependence in hadronic collisions.
The pion asymmetries are compared to two different model predictions, DMP+2013~\cite{ref:D_Alesio_2017} and KPRY~\cite{ref:Kang_2017b}.  The KPRY model provides a better qualitative description of the data, but both model calculations significantly undershoot the observed asymmetries for jet-$p_T > 10$ GeV/$c$. As has been emphasized by both groups, the transverse momentum dependences of the unpolarized and the Collins fragmentation functions are not well understood. In $pp$ collisions, the unpolarized denominator of Eq.~\ref{eq:crosssec} includes a contribution from gluon jets, which has not yet been included in some theoretical calculations of the Collins asymmetry.  This makes data-theory comparisons more difficult to interpret, especially in the low jet-$p_T$, low hadron-$z$ region where gluon jets make a significant contribution. Various two-dimensional presentations of the asymmetries will provide valuable new constraints on the kinematic dependence of the Collins fragmentation function when included in future global analyses.  The measured $K^+$ asymmetries are consistent with those seen for $\pi^+$, while the asymmetries for $K^-$, $p$, and $\bar{p}$ are all consistent with zero.  They provide complementary information regarding the dynamical origins of the Collins fragmentation function.  However, the kaon and proton asymmetries, which are measured here in $pp$ collisions for the first time, are limited by statistics, warranting further investigation with additional data in the future.\par

\begin{acknowledgments}
The authors thank Umberto D’Alesio, Francesco Murgia, Cristian Pisano, Zhong-Bo Kang, Alexei Prokudin, Felix Ringer and Feng Yuan for valuable discussions and providing the results of their model calculations. 

We thank the RHIC Operations Group and RCF at BNL, the NERSC Center at LBNL, and the Open Science Grid consortium for providing resources and support.  This work was supported in part by the Office of Nuclear Physics within the U.S. DOE Office of Science, the U.S. National Science Foundation, National Natural Science Foundation of China, Chinese Academy of Science, the Ministry of Science and Technology of China and the Chinese Ministry of Education, the Higher Education Sprout Project by Ministry of Education at NCKU, the National Research Foundation of Korea, Czech Science Foundation and Ministry of Education, Youth and Sports of the Czech Republic, Hungarian National Research, Development and Innovation Office, New National Excellency Programme of the Hungarian Ministry of Human Capacities, Department of Atomic Energy and Department of Science and Technology of the Government of India, the National Science Centre and WUT ID-UB of Poland, the Ministry of Science, Education and Sports of the Republic of Croatia, German Bundesministerium f\"ur Bildung, Wissenschaft, Forschung and Technologie (BMBF), Helmholtz Association, Ministry of Education, Culture, Sports, Science, and Technology (MEXT) and Japan Society for the Promotion of Science (JSPS).

\end{acknowledgments}

\appendix*

\section{Collins asymmetries for $x_{F} < 0$}

\begin{figure*}
\centering
    \includegraphics[width=2.0\columnwidth]{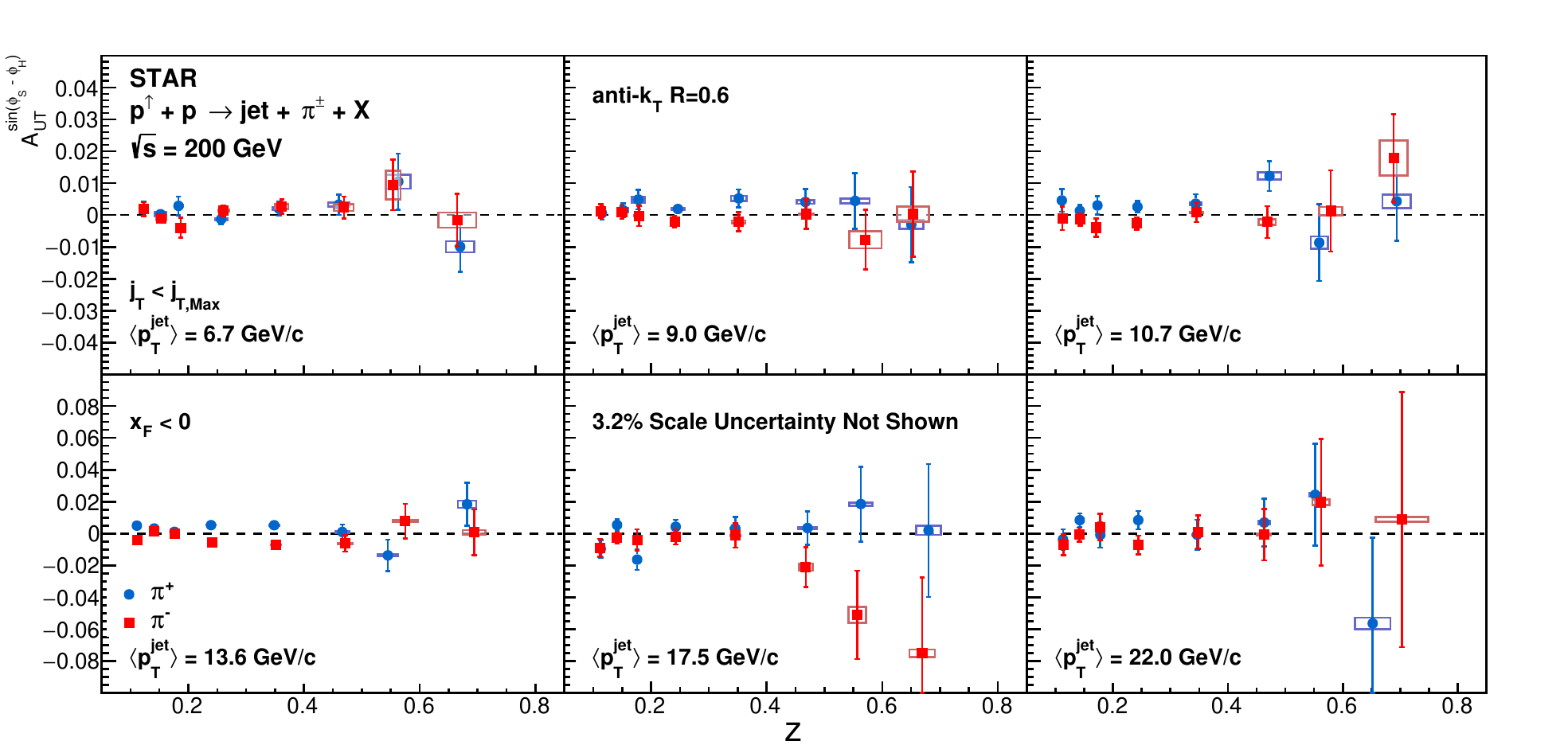}
    \caption{Collins asymmetries, $A_{UT}^{\sin(\phi_{S}-\phi_{H})}$, as a function of the charged pion's longitudinal momentum fraction, $z$, in different jet-$p_{T}$ bins for $x_{F} < 0$. The bars show the statistical uncertainties, while the size of the boxes represents the systematic uncertainties on $A_{UT}^{\sin(\phi_{S}-\phi_{H})}$ (vertical) and hadron-$z$ (horizontal).}
    \label{fig:figure_Collins_vs_z_revEta}
\end{figure*}

\begin{figure*}
\centering
    \includegraphics[width=2.0\columnwidth]{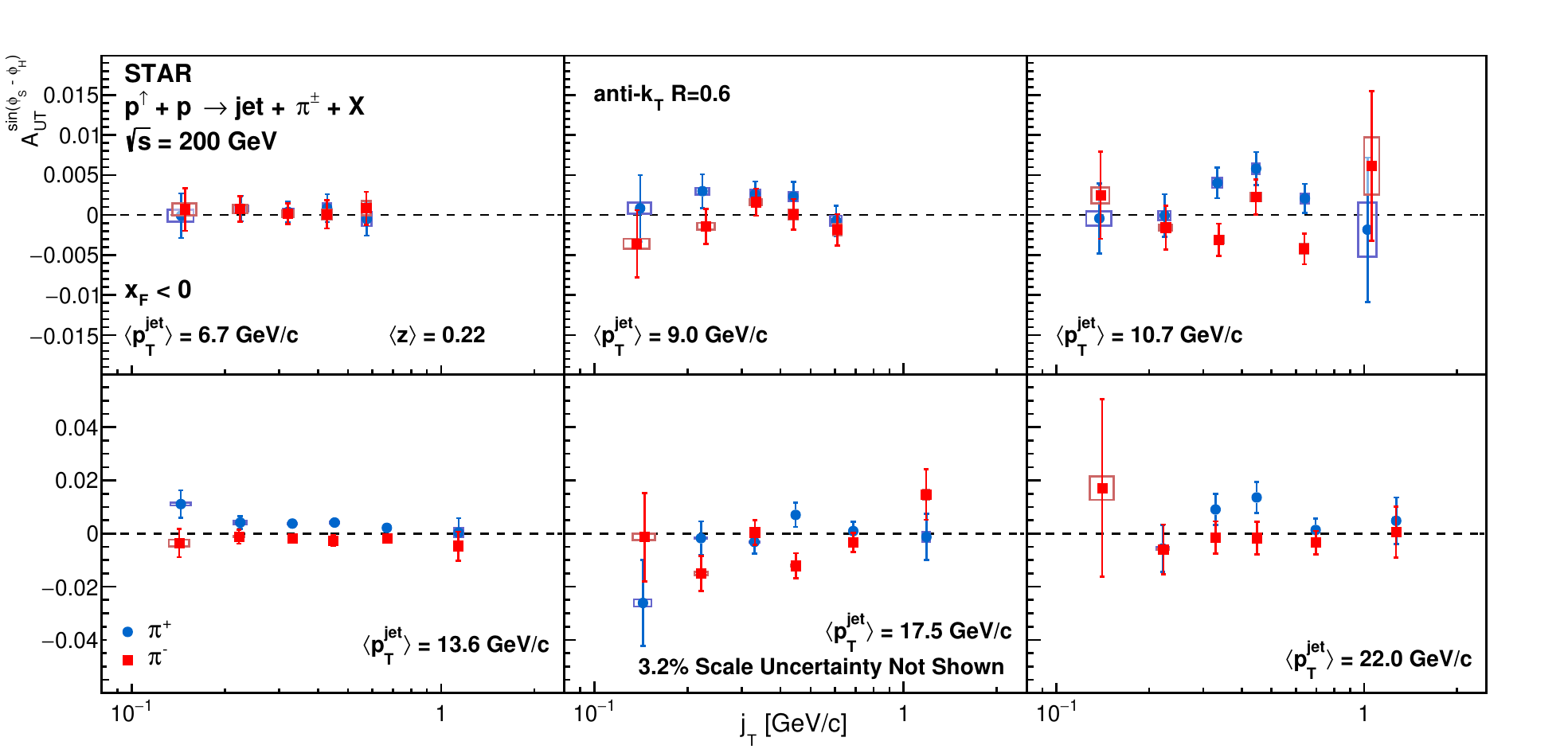}
    \caption{Collins asymmetries, $A_{UT}^{\sin(\phi_{S}-\phi_{H})}$, as a function of the charged pion's momentum transverse to the jet axis, $j_{T}$, in different jet-$p_{T}$ bins for $x_{F} < 0$. The bars show the statistical uncertainties, while the size of the boxes represents the systematic uncertainties on $A_{UT}^{\sin(\phi_{S}-\phi_{H})}$ (vertical) and hadron-$j_{T}$ (horizontal).}
    \label{fig:figure_Collins_vs_jT_jetPt_revEta}
\end{figure*}

\begin{figure}[!hbt]
    \includegraphics[width=1.0\columnwidth]{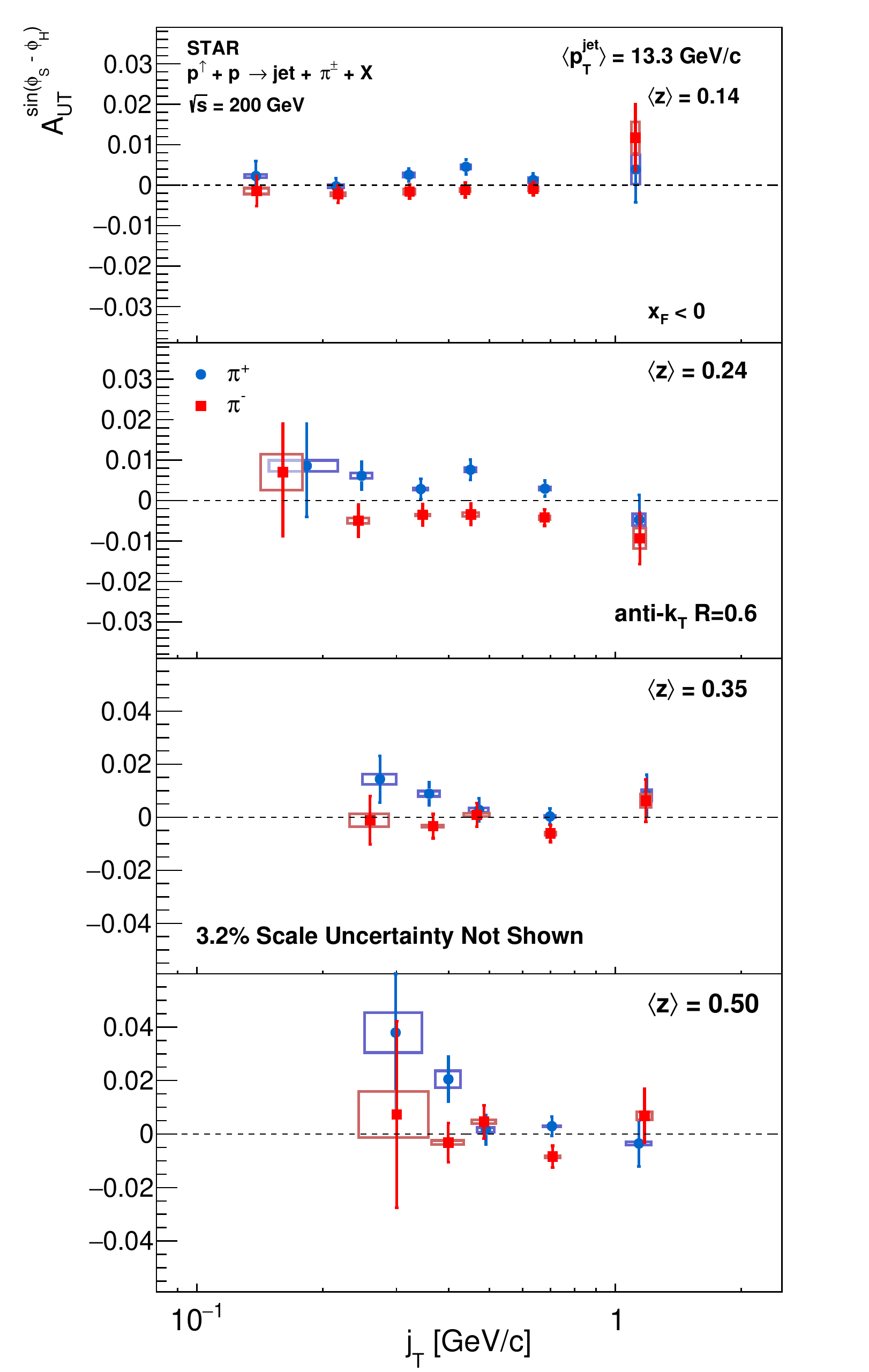}
    \caption{Collins asymmetries, $A_{UT}^{\sin(\phi_{S}-\phi_{H})}$, as a function of the charged pion's momentum transverse to the jet axis, $j_{T}$, in different hadron longitudinal momentum fraction $z$ bins, integrated over detector jet-$p_T > 9.9$ GeV/$c$ for $x_{F} < 0$. The bars show the statistical uncertainties, while the size of the boxes represents the systematic uncertainties on $A_{UT}^{\sin(\phi_{S}-\phi_{H})}$ (vertical) and hadron-$j_{T}$ (horizontal).}
    \label{fig:figure_Collins_vs_jT_zbin_revEta}
\end{figure}

The Collins asymmetries $A_{UT}^{\sin(\phi_{S}-\phi_{H})}$ with multidimensional binning scheme for $x_{F} < 0$ are presented in Figs.~\ref{fig:figure_Collins_vs_z_revEta}, ~\ref{fig:figure_Collins_vs_jT_jetPt_revEta}, ~\ref{fig:figure_Collins_vs_jT_zbin_revEta}.\par

\bibliography{main}

\end{document}

%% file: star-author-list-2022-09-13.aps.tex
\affiliation{American University of Cairo, New Cairo 11835, New Cairo, Egypt}
\affiliation{Texas A\&M University, College Station, Texas 77843}
\affiliation{Brookhaven National Laboratory, Upton, New York 11973}
\affiliation{AGH University of Science and Technology, FPACS, Cracow 30-059, Poland}
\affiliation{Ohio State University, Columbus, Ohio 43210}
\affiliation{University of Kentucky, Lexington, Kentucky 40506-0055}
\affiliation{Panjab University, Chandigarh 160014, India}
\affiliation{Variable Energy Cyclotron Centre, Kolkata 700064, India}
\affiliation{Central China Normal University, Wuhan, Hubei 430079 }
\affiliation{Abilene Christian University, Abilene, Texas   79699}
\affiliation{Instituto de Alta Investigaci\'on, Universidad de Tarapac\'a, Arica 1000000, Chile}
\affiliation{University of California, Riverside, California 92521}
\affiliation{University of Houston, Houston, Texas 77204}
\affiliation{State University of New York, Stony Brook, New York 11794}
\affiliation{University of Jammu, Jammu 180001, India}
\affiliation{Czech Technical University in Prague, FNSPE, Prague 115 19, Czech Republic}
\affiliation{Nuclear Physics Institute of the CAS, Rez 250 68, Czech Republic}
\affiliation{Shanghai Institute of Applied Physics, Chinese Academy of Sciences, Shanghai 201800}
\affiliation{Yale University, New Haven, Connecticut 06520}
\affiliation{University of California, Davis, California 95616}
\affiliation{Lawrence Berkeley National Laboratory, Berkeley, California 94720}
\affiliation{University of California, Los Angeles, California 90095}
\affiliation{Indiana University, Bloomington, Indiana 47408}
\affiliation{Warsaw University of Technology, Warsaw 00-661, Poland}
\affiliation{Shandong University, Qingdao, Shandong 266237}
\affiliation{Fudan University, Shanghai, 200433 }
\affiliation{University of Science and Technology of China, Hefei, Anhui 230026}
\affiliation{Tsinghua University, Beijing 100084}
\affiliation{University of California, Berkeley, California 94720}
\affiliation{ELTE E\"otv\"os Lor\'and University, Budapest, Hungary H-1117}
\affiliation{University of Heidelberg, Heidelberg 69120, Germany }
\affiliation{Wayne State University, Detroit, Michigan 48201}
\affiliation{Indian Institute of Science Education and Research (IISER), Berhampur 760010 , India}
\affiliation{Kent State University, Kent, Ohio 44242}
\affiliation{Rice University, Houston, Texas 77251}
\affiliation{University of Tsukuba, Tsukuba, Ibaraki 305-8571, Japan}
\affiliation{University of Illinois at Chicago, Chicago, Illinois 60607}
\affiliation{Lehigh University, Bethlehem, Pennsylvania 18015}
\affiliation{University of Calabria \& INFN-Cosenza, Italy}
\affiliation{National Cheng Kung University, Tainan 70101 }
\affiliation{Purdue University, West Lafayette, Indiana 47907}
\affiliation{Southern Connecticut State University, New Haven, Connecticut 06515}
\affiliation{Technische Universit\"at Darmstadt, Darmstadt 64289, Germany}
\affiliation{Temple University, Philadelphia, Pennsylvania 19122}
\affiliation{Valparaiso University, Valparaiso, Indiana 46383}
\affiliation{Indian Institute of Science Education and Research (IISER) Tirupati, Tirupati 517507, India}
\affiliation{Institute of Modern Physics, Chinese Academy of Sciences, Lanzhou, Gansu 730000 }
\affiliation{Frankfurt Institute for Advanced Studies FIAS, Frankfurt 60438, Germany}
\affiliation{National Institute of Science Education and Research, HBNI, Jatni 752050, India}
\affiliation{University of Texas, Austin, Texas 78712}
\affiliation{Rutgers University, Piscataway, New Jersey 08854}
\affiliation{Institute of Nuclear Physics PAN, Cracow 31-342, Poland}
\affiliation{Max-Planck-Institut f\"ur Physik, Munich 80805, Germany}
\affiliation{Creighton University, Omaha, Nebraska 68178}
\affiliation{Indian Institute Technology, Patna, Bihar 801106, India}
\affiliation{Ball State University, Muncie, Indiana, 47306}
\affiliation{Universidade de S\~ao Paulo, S\~ao Paulo, Brazil 05314-970}
\affiliation{Huzhou University, Huzhou, Zhejiang  313000}
\affiliation{Michigan State University, East Lansing, Michigan 48824}
\affiliation{Argonne National Laboratory, Argonne, Illinois 60439}
\affiliation{United States Naval Academy, Annapolis, Maryland 21402}
\affiliation{South China Normal University, Guangzhou, Guangdong 510631}

\author{M.~S.~Abdallah}\affiliation{American University of Cairo, New Cairo 11835, New Cairo, Egypt}
\author{B.~E.~Aboona}\affiliation{Texas A\&M University, College Station, Texas 77843}
\author{J.~Adam}\affiliation{Brookhaven National Laboratory, Upton, New York 11973}
\author{L.~Adamczyk}\affiliation{AGH University of Science and Technology, FPACS, Cracow 30-059, Poland}
\author{J.~R.~Adams}\affiliation{Ohio State University, Columbus, Ohio 43210}
\author{J.~K.~Adkins}\affiliation{University of Kentucky, Lexington, Kentucky 40506-0055}
\author{I.~Aggarwal}\affiliation{Panjab University, Chandigarh 160014, India}
\author{M.~M.~Aggarwal}\affiliation{Panjab University, Chandigarh 160014, India}
\author{Z.~Ahammed}\affiliation{Variable Energy Cyclotron Centre, Kolkata 700064, India}
\author{D.~M.~Anderson}\affiliation{Texas A\&M University, College Station, Texas 77843}
% ...

\collaboration{STAR Collaboration}\noaffiliation